\documentclass[%
 pre,reprint,
 superscriptaddress,
 amsmath,amssymb,
 aps,
]{revtex4-2}

\usepackage{graphicx}
\usepackage{subcaption}
\usepackage[font={small}]{caption}
\usepackage{dcolumn}
\usepackage{bm}
\usepackage[hidelinks]{hyperref}
\usepackage{comment}

\usepackage{xcolor}

\begin{document}

\preprint{APS/123-QED}

\title{Densely packed particle raft at vertically vibrated air-water interface}

\author{Xiuhe Yan}
\thanks{These two authors contributed equally to the work.}
\author{Tabitha C. Watson}
\thanks{These two authors contributed equally to the work.}
\author{Hongyi Xiao}%
 \email{hongyix@umich.edu}
\affiliation{%
 Department of Mechanical Engineering, University of Michigan, Ann Arbor, USA
}%

\date{\today}

\begin{abstract}
We investigate the dynamics of a dense raft of millimeter-sized granular particles at a vertically vibrated air-water interface, which displays a rich set of patterns and particle dynamics as we vary the vibration amplitude, frequency, and particle packing fraction. 
While the classical parametric instability with standing waves still occurs over a certain range of parameters, the measured wave dispersion relations indicate an increasing role of the raft's emergent elasticity at higher packing fractions, where the effective surface tension decreases and the out-of-plane bending modulus increases.
At higher vibration frequencies and lower amplitudes, we identify a regime without standing waves. Instead, individual particles exhibit thermal-like motion, with transport crossing over from diffusive to sub-diffusive as the packing fraction increases. The particle dynamics also display spatial and temporal heterogeneity, as in supercooled liquids.  
Starting from this regime, when the vibration amplitude is further increased, a large cavity eventually forms inside the raft, whose size and shape depend on the vibration frequency and the injected vibration energy. The cavitation results in the coexistence of free-surface water waves inside the cavity and thermal-like particle motion in the surrounding raft.
\end{abstract}

\maketitle

\section{\label{sec:intro} Introduction}

Particles residing at an interface between two fluids can assemble into a raft via attractive capillary interactions~\cite{protiere2023particle}, which arise from the energy minimization of the distorted liquid interface~\cite{nicolson1949interaction,kralchevsky2000capillary,vella2005cheerios}. 
Particle rafts have wide applications such as self-assembly, interfacial stabilization, and pollution control~\cite{protiere2023particle}. Because particles are macroscopic and directly observable, the rafts they form are also ideal model systems for understanding the packing structure and dynamics of out-of-equilibrium systems like athermal amorphous solids~\cite{xiao2020strain,xiao2023identifying,hobson2026structural}. Particle rafts also share similarities with living matter at interfaces, such as fish bubble nests~\cite{hostache1998reproductive} and rafts of mosquito eggs, fire ants, and starfish embryos~\cite{mlot2011fire,tan2022odd}, where internal activity and environmental fluctuations drive complex dynamics and pattern formation.
Yet, even the response of passive rafts to dynamical excitations is poorly understood because of the complex and many-body nature of the capillary and hydrodynamic interactions~\cite{dalbe2011aggregation,dani2015hydrodynamics}. 

Particles at a quiescent fluid interface often pack densely and collectively exhibit solid-like features~\cite{protiere2023particle}. 
Several studies have examined the mechanical failure of particle rafts via imposing tensile strains~\cite{kim2019failure,xiao2020strain,to2023rifts,xiao2023identifying,hobson2026structural}, rotary shear~\cite{sasaki2026constitutiveflowlawhydrogel}, local intrusion~\cite{kozlowski2019dynamics}, and surfactant injection~\cite{Vella2004212,peco2017surfacetension}.
In these tests, the rafts often demonstrate a certain degree of elasticity before failure via fracturing~\cite{Vella2004212,peco2017surfacetension,to2023rifts} or shear banding~\cite{xiao2020strain,hobson2026structural,sasaki2026constitutiveflowlawhydrogel}.
Moreover, the deformation of particle rafts can be complicated by out-of-plane motion induced by compression and indentation~\cite{Vella2004212,Aveyard20001969,aveyard2000structure,vella2006dynamics,zang2010viscoelastic}, which can result in the raft buckling and collapsing~\cite{abkarian2013gravity,protiere2017sinking,druecke2023collapse,sayyari2025destabilizing}. 
In this case, a bending modulus arises, resisting the change in the interface energy associated with its curvature, since the liquid-particle contact angle needs to be maintained~\cite{kralchevsky2005,aveyard2003,danov2001capillary}.
While these solid-like aspects of particle rafts are well documented, their liquid-like behaviors remain less explored (except for like-charged colloids). Since interfacial particle rearrangements are hindered by attractive interactions and by strong dissipation from friction and viscosity, rafts seldom access liquid-like regimes, which calls for experiments with more dynamic excitations.

Under dynamic excitation, interparticle hydrodynamic interactions and underlying fluid flow can become important in addition to capillarity~\cite{planchette2012surface, saddier2024breaking,Pak1993,Lagarde2020}.
Surface waves generated by vibration can effectively excite the particles at the interface. 
While several studies have reported the dynamics of floating particles at vibrated interfaces in the limit of low packing fraction (e.g., rotation and synchronization)~\cite{Barotta2023,Sungar2025,Barotta2025,barotta2025macroscopicbrownianmotionchaotic}, only a handful of studies have examined densely packed particle rafts at vibrated interfaces with propagating surface waves~\cite{planchette2012surface,van2020propagation,saddier2024breaking}.
For example, important raft properties such as the bending modulus can be probed from the observed wave dispersion~\cite{planchette2012surface,van2020propagation}.
Propagating surface waves can also trigger the fragmentation of a raft, resulting in a floe-size distribution similar to that of floating sea ice~\cite{saddier2024breaking}. 
This is related to viscous stresses from the underlying fluid, a mechanism that is absent in quasi-static processes.
However, in these studies, the waves were generated by a localized source placed directly at the free surface, e.g., a vertically vibrating plate, making the raft dynamics intrinsically heterogeneous. 
To impose homogeneous and dynamic excitation, we propose to vertically shake the entire liquid container, e.g., triggering Faraday waves~\cite{faraday1830peculiar}.
Still, the corresponding collective particle dynamics and emergent pattern formation remain largely unexplored.

Under periodic vertical oscillations, a liquid interface can exhibit Faraday waves when the vibration amplitude exceeds a critical threshold~\cite{edwards1994patterns}. This originates from the parametric Faraday instability in which standing waves emerge and oscillate typically at half the driving frequency~\cite{benjamin1954instability,kumar1994parametric,zhang1997pattern}. The wave formation depends on several parameters, such as the excitation frequency, vibration amplitude, the container geometry, the fluid depth, and the fluid’s physical properties including density, viscosity, and surface tension~\cite{douady1990experimental,edwards1994patterns,lioubashevski1997scaling,batson2013parametric}. Higher excitation frequencies generally lead to shorter wavelengths and higher threshold accelerations, whereas increasing the fluid viscosity or surface tension tends to stabilize the surface against wave formation. 
By varying these control parameters, various spatial patterns can be obtained, such as stripes, squares, and hexagons, corresponding to different modes of parametric resonance~\cite{miles1993faraday,chen1997pattern,zhang2023pattern,shao2021surface}.
At higher vibration amplitudes, the standing-wave patterns can lose stability through secondary instabilities~\cite{ezerskii1985random,zhang1995secondary} and transition to spatiotemporal chaos~\cite{ezersii1986spatiotemporal,tufillaro1989order,bosch1993spatiotemporal,frumkin2023coupled}, reflecting non-linear system dynamics~\cite{zhang1997pattern,shani2010localized}.

Interestingly, Faraday's original 1831 study of vibrated fluid interfaces~\cite{faraday1830peculiar} was attached to a manuscript discussing patterns formed by a layer of particles on a vibrating plate, which also shows rich phenomena arising from energy transfer via inelastic particle collisions~\cite{Pak1993,melo1994transition,umbanhowar1996localized,rouyer2000velocity,feitosa2002breakdown,deseigne2010collective,scholz2018rotating,zheng2026topological}, such as the formation of subharmonic standing waves~\cite{melo1994transition} and energy localization~\cite{umbanhowar1996localized}. Here, we combine both systems and consider a dense raft of granular particles on a vertically vibrated air-water interface.
While the dynamics of Faraday waves on a liquid surface can be described by the incompressible Navier--Stokes equations with associated kinematic boundary conditions~\cite{benjamin1954instability,miles1993faraday,kumar1994parametric}, the addition of a dense layer of particles can induce additional dissipation and elasticity in the interface. This setup then leads to two coupled questions: how densely packed particles influence the wave patterns, and how vertical vibration excites particle dynamics, which can be intriguing when the supposed wavelength approaches the particle size.

In this study, we experimentally investigate the pattern formation and particle dynamics of dense granular rafts at a vertically vibrated air-water interface. 
By systematically varying the particle packing fraction, vibration frequency, and amplitude, we identify distinct regimes of raft behaviors including not only the formation of standing waves, but also a glassy regime in which waves disappear and particles exhibit thermal-like dynamics, as well as a regime in which a large cavity in the raft develops.
We then perform quantitative measurements to understand how particle-particle and particle-fluid interactions influence these observed phenomena. Aside from relevance to real-world systems, the rich physics revealed in our experiments demonstrates that particle rafts can serve as model systems for understanding the implications of complex particle interactions for a variety of problems beyond the elastoplastic behavior of solids.

\section{Experimental methods}

\begin{figure}[t]
    \centerline{\includegraphics[width=0.95\linewidth]{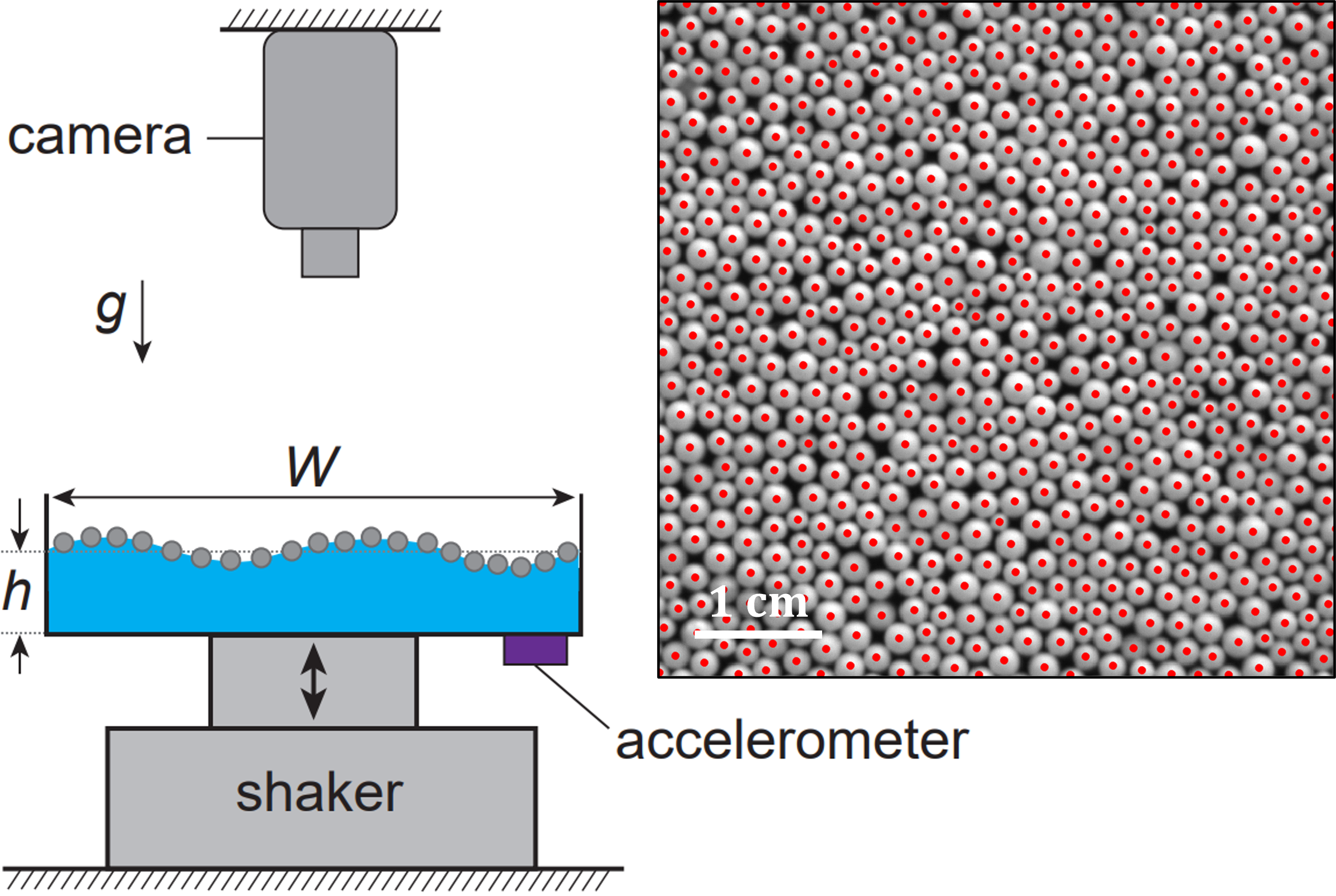}}
    \caption{\label{fig:setup} Schematic of the experiment with a vibrated particle raft and an example snapshot with tracked particle centers labeled as red dots. }
\end{figure}

We constructed an experimental apparatus to vertically vibrate a dense granular raft, as shown in the schematic in Fig.~\ref{fig:setup}.
The container has a square cross-section with side length $W=10.9$\,cm and is filled with deionized water to a depth of $h=5$\,mm. 
Driven by a shaker, it experiences a sinusoidal oscillation with its vertical position following $Z(t)=A\sin(2\pi ft)$, where $f$ is the driving frequency and $A$ is the vibration amplitude.
We used an accelerometer to monitor the container's acceleration amplitude $a$, and calculated the displacement amplitude via $A=a/(4\pi^2f^2)$. 
We tested various frequencies in a range $f\in[20,95]$\,Hz and amplitudes in $A\in[0.02,0.2]$\,mm. 

The particles are closed-cell Styrofoam spheres with a density of $\rho_p\approx 15$ $\mathrm{kg}\,\mathrm{m}^{-3}$, which is small compared to the nominal density of water, $\rho=1000$ $\mathrm{kg}\,\mathrm{m}^{-3}$. 
Two batches of size-polydisperse particles were tested with different mean diameters, $d = 1.4\pm0.2$\,mm and $d = 2.1 \pm0.2$\,mm. 
While the small particles are useful for studying pattern formation, the large particles present an easier case for tracking individual particles and resolving their dynamics. 
We employed a sessile drop technique to determine the three-phase contact angle for our particles at the air-water interface~\cite{extrand2008}. Results from tests with different particle diameters and drop volumes indicate the angle to be $87^{\circ} \pm 6^{\circ}$, consistent with those reported for bulk polystyrene~\cite{li2007, kwok1998}.
Prior to each experiment, we deposited particles on the water surface and removed any stacked particles to obtain a strict monolayer. We define the particle packing fraction $\phi$ as the ratio between the projected area occupied by the particles and the cross-sectional area of the container.  

\begin{figure*}[t!]
  \centering
  \includegraphics[width=0.98\linewidth]{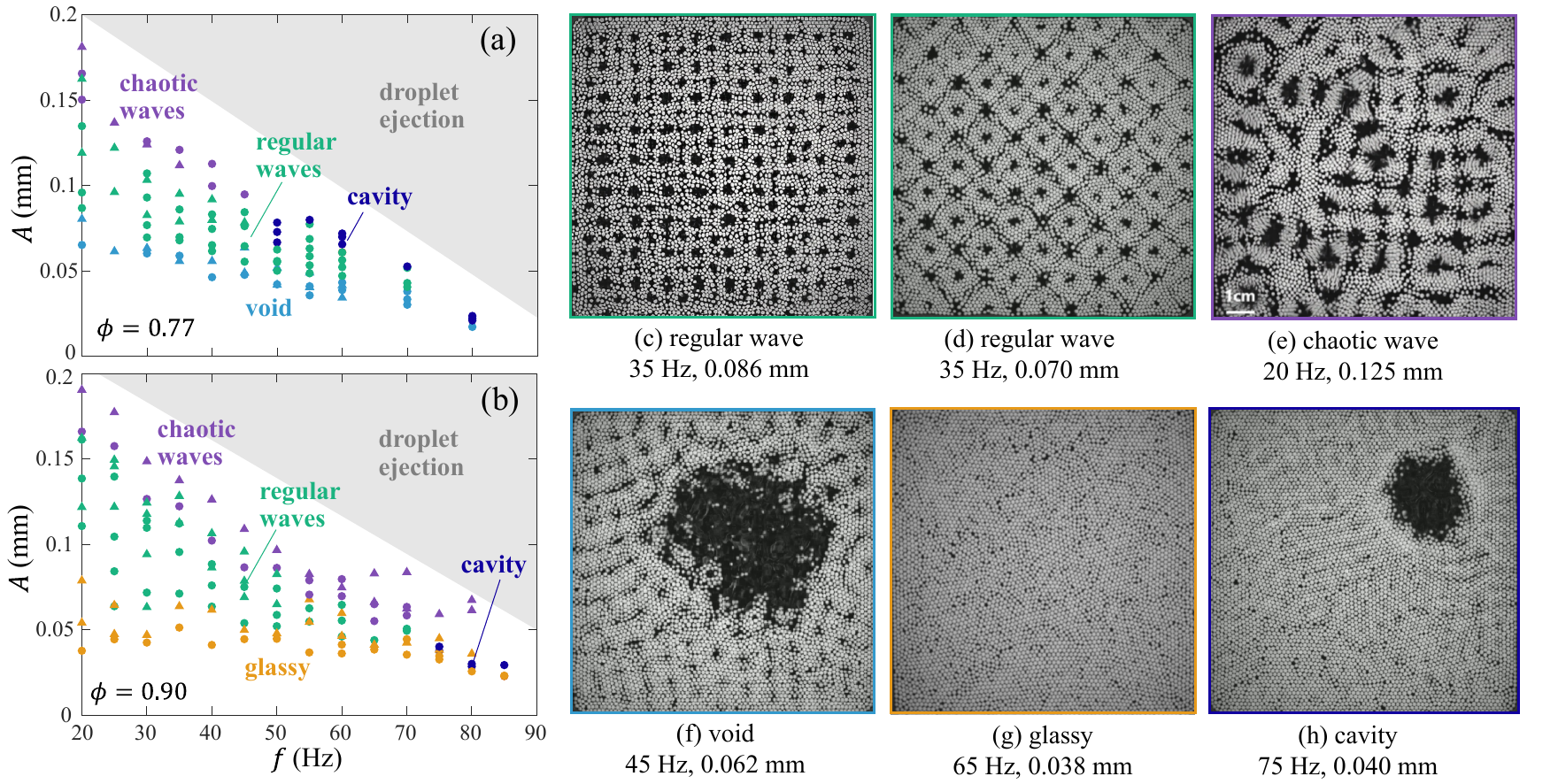}
  \caption{Phase diagram at packing fractions (a) $\phi=0.77$ and (b) $\phi=0.90$. The circles represent the data for particles with $d=1.4$\,mm, and the triangles for $d=2.1$\,mm. The shaded region in each panel approximately indicates a regime with water droplets being ejected. (c)-(h) Experimental snapshots for the respective regimes. Specifically, (e) and (f) are from $\phi=0.77$ and (c), (d), (g), and (h) are from $\phi=0.90$. For each snapshot, see the corresponding video in the supplementary~\cite{supp}.}
  \label{fig:phase}
\end{figure*}

For observing pattern formation, images with a resolution of 2448$\times$2048\,pixels were recorded by a camera at 73\,Hz.  
Particle positions were extracted using a standard phase-coding circular Hough transform method 
(available in MATLAB) to detect centers of the circular projections of the particles~\cite{atherton1999detection}, with an example result shown in Fig.~\ref{fig:setup}. 
To estimate the packing fraction $\phi$, we binarized an image of the initial packing configuration without vibration and calculated $\phi$ based on the number of pixels belonging to the particles. Varying the binarization threshold within a reasonable range may influence the obtained packing fraction systematically up to 2.4\%. 
To obtain the intra-cycle particle dynamics, we used a high-speed camera to monitor a few selected cases, which captured images at 460\,Hz with a resolution of 2448$\times$1024\,pixels.

\section{Pattern formation in vibrated rafts}
\label{sec:phase}

Under different vibration amplitudes $A$, driving frequencies $f$, and particle packing fractions $\phi$, a raft exhibits distinct dynamics and patterns, which are summarized in Fig.~\ref{fig:phase}a,b for $\phi=0.77$ and $\phi=0.90$, respectively. Both results for $d=1.4$\,mm and $d=2.1$\,mm are displayed, showing similar general behaviors, which we categorize into the following five regimes.

\begin{figure}[h!]
    \centering \includegraphics[width=1\linewidth]{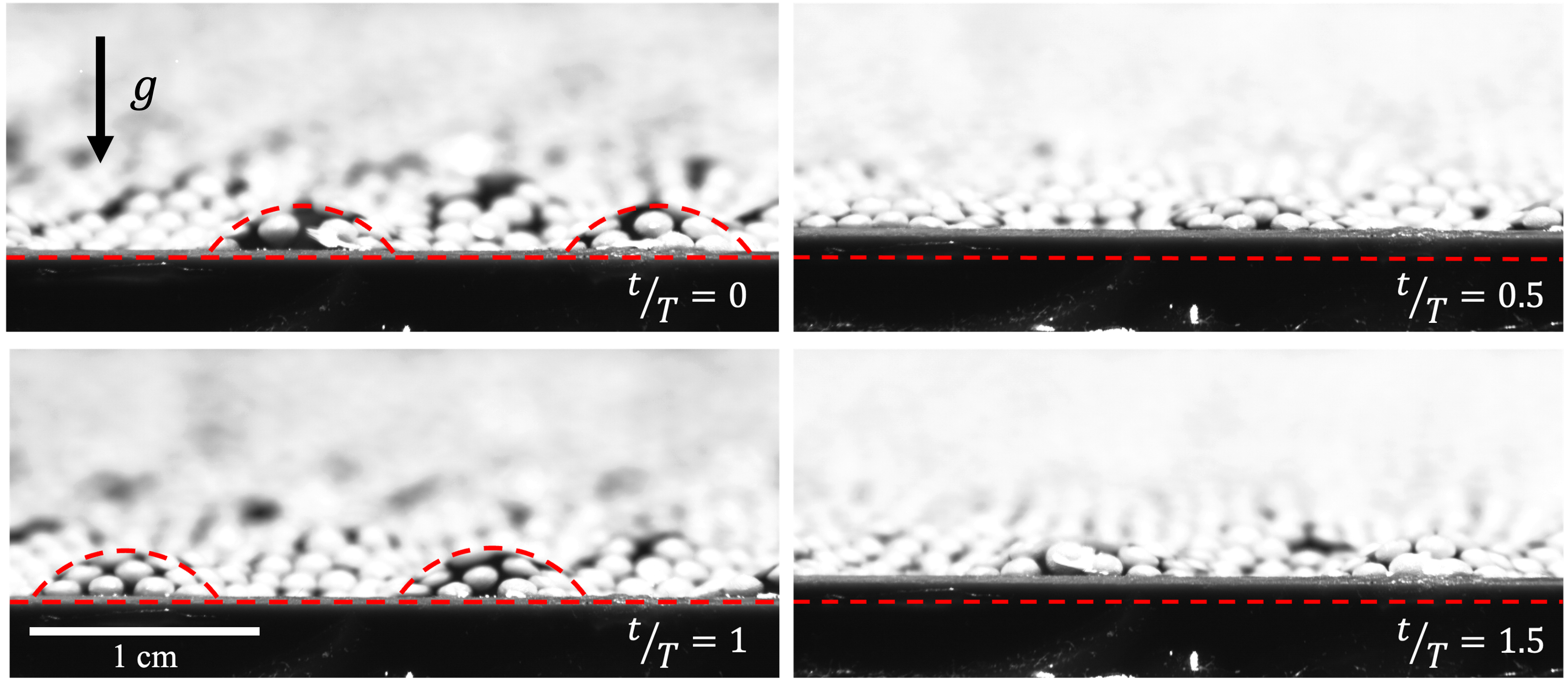}
    \caption{Time series of a vibrated raft of $d=1.4$\,mm particles in the standing-wave regime, $A=0.086$\,mm, $f=35$\,Hz, demonstrating that the wave frequency is half of the driving frequency. See the corresponding video in the supplementary~\cite{supp}.}
    \label{fig:freqcomp}
\end{figure}

We first identify a regime with regular patterns formed by standing waves, such as the square-shaped patterns shown in Fig.~\ref{fig:phase}c and d, which closely resemble the Faraday wave patterns of a pure liquid system~\cite{milner1991square,miles1994faraday}. Particles near the wave antinodes pack loosely, making the periodicity obvious. 
Other patterns also exist in this regime, such as striped patterns which have only one family of waves along a single direction. 
In Fig.~\ref{fig:freqcomp}, we show side views of the raft over two vibration cycles, with time normalized by $T=1/f$. 
At $t/T=0.5$ and $t/T=1.5$ no peaks are observable, corresponding to the time instances in which the interface is flat.
At both $t/T=0$ and $t/T=1$, two sets of wave peaks are observed, but occur at different locations offset by half the wavelength, which indicates that the standing waves have a period of $2T$. Hence, they oscillate at half the driving frequency like classical Faraday waves due to parametric resonance~\cite{miles1993faraday,kumar1994parametric,benjamin1954instability}. 
These results suggest that, in this regime, the presence of particles does not fundamentally alter the classical physics of the Faraday instability. The influence of the particles will be discussed in Sec.~\ref{sec:dispersion}.

At small to intermediate frequencies, a transition to a chaotic regime occurs for both $\phi=0.77$ and $\phi=0.90$ as we increase the amplitude $A$. The wave patterns are no longer regular and evolve over time, see a snapshot in Fig.~\ref{fig:phase}e, and the transition amplitude is lower with increasing frequency. 
In pure liquid systems, the transition from spatially ordered, stable patterns to spatiotemporally chaotic patterns can occur via different pathways, depending on the system configuration~\cite{ezersii1986spatiotemporal,tufillaro1989order,bosch1993spatiotemporal,ghadiri2021control,frumkin2023coupled}. An essential element in such transitions is the development of defects within the regular patterns~\cite{tufillaro1989order,shani2010localized}; these defects arise from secondary instabilities of the ordered standing waves, such as those driven by non-linear mode coupling~\cite{milner1991square,zhang1997pattern} or localized phase instabilities~\cite{shani2010localized}.
In our case, the presence of particles in the regular wave regime may already distort the standing waves, since particles do not pack in perfectly periodic configurations. This may open additional pathways for triggering the transition to chaos that are not present in pure liquid systems.

The following three regimes to be discussed have no counterpart phenomena in a pure liquid system. 
If we decrease the amplitude from that of the regular pattern-forming regime, the particles are less agitated and pack closely with their neighbors.
At the lower packing fraction that was tested, $\phi=0.77$, there are not enough particles to cover the entire area, leading to a void that forms in the raft and coexists with standing waves, see Fig.~\ref{fig:phase}f. 

At the higher packing fraction, $\phi=0.90$, the raft behavior is qualitatively different at lower amplitude, $A\approx0.05$\,mm, and exhibits neither standing waves nor a void. Instead, a relatively uniform packing structure is observed and individual particles vibrate with a thermal-like motion, see example in Fig.~\ref{fig:phase}g. The packing is disordered due to the polydispersity of the particles. As discussed later in Sec.~\ref{sec:dense}, particles may be confined by their neighbors and show a sub-diffusive and glassy behavior.

At higher frequencies, $f\ge75$\,Hz, a more peculiar regime emerges, characterized by a cavity forming in the raft, creating a highly dense layer of particles around it with the rest of the raft remaining relatively densely packed, see Fig.~\ref{fig:phase}h. 
Here, we use the term cavity to distinguish it from the void formed at the lower density due to having an insufficient number of particles to cover the liquid surface.  
Still, for $\phi=0.77$, a similar phenomenon is also observed at amplitudes above the regular pattern regime. It is labeled as cavity-forming in Fig.~\ref{fig:phase}a, although merging this regime with the void-forming regime can also be reasonable. 

Lastly, we note that results with both high frequency and high amplitude are not included, as water droplets and particles may be ejected due to the high energy input, making the experiments unsustainable. This is indicated in gray in Fig.~\ref{fig:phase}a,b. That said, this wave-breaking phenomenon~\cite{goodridge1997viscous,goodridge1999breaking} also occurs on the free liquid surface in the cavity-forming regime (see Fig.~\ref{fig:cavity}b).
As the observed raft behaviors are rich, understanding how the particles influence the formation, stability, and transitions of all the observed regimes is a complex task beyond the scope of this work, especially for the transient behavior in the chaotic regime. 
We will focus on characterizing the steady-state behaviors of the regular wave regime, the glassy regime, and the cavity-forming regime.

\section{Dispersion relations in regular wave patterns}
\label{sec:dispersion}

In the regular wave regime, we investigate how the presence of particles influences the dispersion relation, which links the wave's oscillation frequency and spatial periodicity. 
For inviscid and incompressible fluids, a classical dispersion relation is  
\begin{equation}
\omega^2 = \left( gk + \frac{\gamma}{\rho}k^3 \right) \tanh(kh),
\label{eq:dispersion_classic}
\end{equation}
where $\omega$ is the angular frequency of the surface wave, $k$ is the wave number, $g$ is the gravitational acceleration, and $\gamma$ is the surface tension, which is 72\,$\mathrm{mN}\,\mathrm{m}^{-1}$ for water.
The term $\tanh(kh)$ accounts for a finite fluid depth.
The terms $gk$ and $(\gamma/\rho)k^3$ represent the gravitational and capillary restoring forces, which dominate at long and short wavelengths, respectively.  
Setting these two terms equal yields the capillary length, $1/k=\sqrt{\gamma/g\rho}$, which is about 2.7 mm for water, corresponding to $k\approx370\,\mathrm{rad}\,\mathrm{m}^{-1}$ and the wavelength $\lambda=2\pi/k\approx1.7$\,cm. 
This is on the same order of magnitude as the wavelengths in Figs.~\ref{fig:phase} and  \ref{fig:freqcomp}, indicating that both effects are at play.

For a particle raft, having densely packed particles may alter both the effective surface tension and the inertia of the interface. 
At high packing fractions, the particles may repel each other when they are in contact, resulting in a reduced surface tension, $\gamma_e$, which is referred to as the effective stretching modulus~\cite{planchette2012surface}. 
Moreover, as standing waves distort the curvature of the liquid surface between particles, a bending modulus, $B$, emerges as the raft resists out-of-plane deformation.
The dispersion relation can then be modified as
\begin{equation}
\omega^2 \left(1 + \frac{\rho_r}{\rho} a_p k \right)
= g k \left( 1 + \frac{\gamma_e k^2}{\rho g} + \frac{B k^4}{\rho g} \right)\tanh(kh),
\label{eq:dispersion_modified}
\end{equation}
where the $k^5$ term accounts for out-of-plane bending by treating the raft as a thin elastic sheet~\cite{landau2012theory}. An additional factor on the left-hand side is included to consider the added particle inertia, in which $\rho_r=\rho_p\phi$ and $a_p=2d/3$ are the effective density and thickness of the raft~\cite{planchette2012surface}, respectively. As $\rho_p\ll\rho$ for Styrofoam particles, the added inertia is small. 
This formulation was introduced to model waves in a particle raft induced by a vibrating plate at the interface~\cite{planchette2012surface}; it may also be applicable to our experiments.

\begin{figure}[h]
    \centering
    \includegraphics[width=0.95\linewidth]{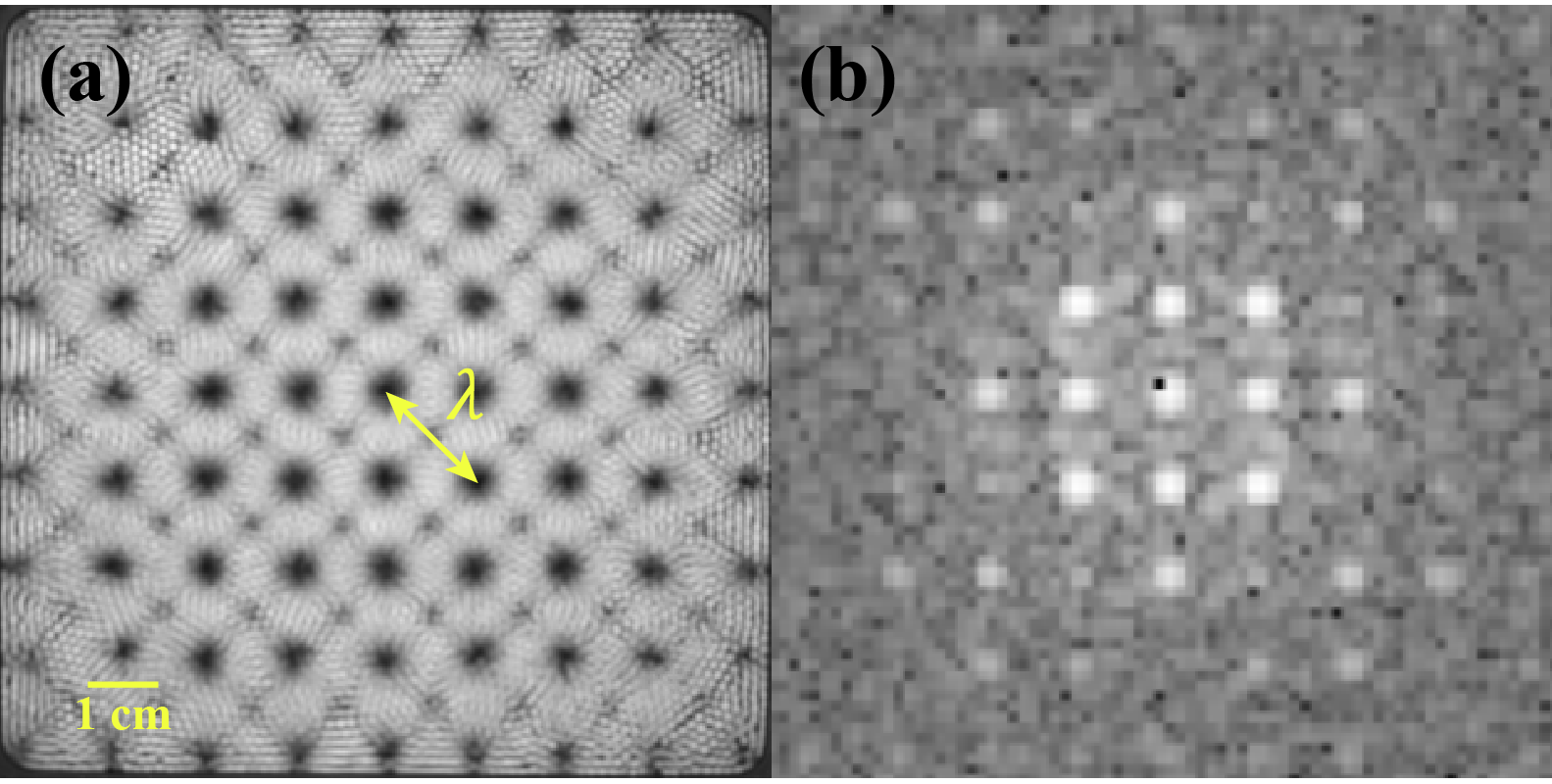}
    \caption{(a) A time-averaged experimental image in the regular wave regime for $d=1.4$\,mm, $f=20$\,Hz, $A=0.124$\,mm, and $\phi=0.81$. (b) The corresponding two-dimensional Fourier transform result, indicating a wavelength of $\lambda=1.72$\,cm.}
    \label{fig:fft}
\end{figure}

We focused on the square-shaped wave patterns and tested five different packing fractions $\phi$ for the smaller particles with $d=1.4$\,mm. At each $\phi$, we extracted a dispersion relation by testing various vibration amplitudes and frequencies.
In each experiment, we performed a time-average for all frames and obtained a single image to capture the periodicity, see Fig.~\ref{fig:fft}a.
A two-dimensional Fourier transform was then applied to the image to determine the dominant wave number $k$, as demonstrated in Fig.~\ref{fig:fft}b. 
As shown earlier in Fig.~\ref{fig:freqcomp}, the corresponding angular frequency is $\omega = \pi f$.

\begin{figure}[h]
    \centering
    \includegraphics[width=1.025\linewidth]{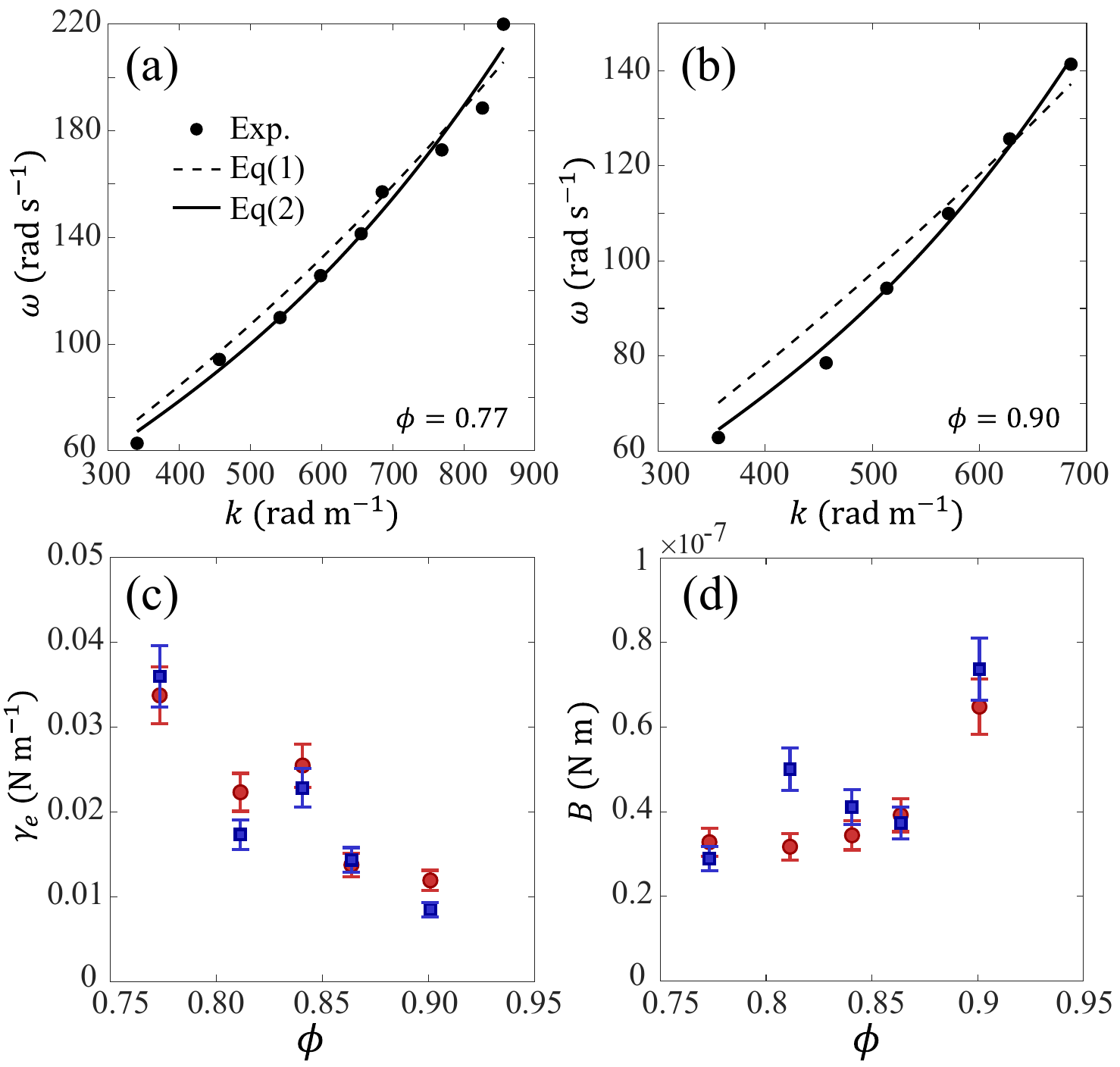}
    \caption{Regular standing wave patterns for $d=1.4$\,mm. (a) Experimental result of angular frequency vs. wave number (circles) for $\phi = 0.77$. The dashed curve represents the fitted Eq.~\ref{eq:dispersion_classic}, with $R^2 = 0.97$, and the solid curve is fit to Eq.~\ref{eq:dispersion_modified}, with $R^2 = 0.98$.
    (b) Experimental result (circle) for $\phi = 0.90$, Eq.~\ref{eq:dispersion_classic} (dashed curve) with $R^2 = 0.95$, and Eq.~\ref{eq:dispersion_modified}, with $R^2 = 0.99$.
    (c) and (d) are the fitted stretching modulus $\gamma_e$ and bending modulus $B$ at various packing fractions $\phi$, respectively.  Red circles are results from linear fitting and blue squares are from non-linear fitting to Eq.~\ref{eq:dispersion_modified}.
    Error bars represent one standard deviation of the uncertainty estimated from fitting residuals.  }
    \label{fig:dispersion}
\end{figure}

Our measured dispersion relations indicate that $k$ increases with $\omega$ following a slightly non-linear trend, see two examples at different packing fractions in Fig.~\ref{fig:dispersion}a and b. For $\phi=0.77$, both Eqs.~\ref{eq:dispersion_classic}~and~\ref{eq:dispersion_modified} can be well-fitted to the measurements, with the coefficients of determination $R^2>0.95$. However, for a higher packing fraction, $\phi = 0.90$, the modified relation, Eq.~\ref{eq:dispersion_modified}, better captures the non-linearity in the dispersion relation, indicating a stronger presence of the bending rigidity via the $Bk^5$ term. 

We next examine the fitted stretching and bending moduli as functions of the packing fraction, see Fig.~\ref{fig:dispersion}c and d. To test the sensitivity of the results to the fitting method for Eq.~\ref{eq:dispersion_modified}, we used both direct non-linear least-squares fitting for the relation between $\omega$ and $k$ and an alternative linear  least-squares fitting, in which we treat the entire left-hand side (minus $gk\tanh(kh)$) as the dependent variable and the two non-linear terms associated with $\gamma_e$ and $B$ on the right-hand side as independent variables, e.g., $\tanh{(kh)}k^5/\rho$. In our case, the factor $\tanh{(kh)}\approx1$, while the added inertial term $\rho_ra_pk/\rho\ll1$, meaning they are not influential to the result. The two fitting methods yield similar values of $\gamma_e$ and $B$, with an anomaly for $B$ at $\phi\approx0.81$, which is likely due to experimental noise.

The effective surface tension $\gamma_e$ exhibits a monotonic decrease for higher packing fraction $\phi$, see Fig.~\ref{fig:dispersion}c. This is likely due to the particles replacing and distorting the free liquid–air interface, while developing repulsive steric contact at higher $\phi$.
The bending modulus $B$ increases slowly with $\phi$ at low-to-moderate $\phi$ but rises rapidly toward the highest tested value at $\phi=0.90$, see Fig.~\ref{fig:dispersion}d.
The rapid increase occurs possibly because, as $\phi$ increases, particles become significantly caged by their neighbors and a persistent contact force network develops, including both elastic and frictional particle contact forces, which may give rise to additional bending rigidity.
The measured dependence of $\gamma_e$ and $B$ on $\phi$ qualitatively agrees with prior experimental measurements in rafts under other forms of excitations~\cite{aveyard2000structure, ap2006properties, planchette2012surface}. However, given that the specific particle packing structure in our system is organized by the standing waves, it is likely that the structure-property relation is unique for the emerging elastic-like behavior of the vertically vibrated raft.

\section{Glassy particle dynamics in dense packings }
\label{sec:dense}

In the regime with glassy particle dynamics, no visible standing waves exist and individual particles vibrate in cages formed by their neighbors over short times while rearranging their packing structure over long times, resembling the dynamics of particles in a supercooled liquid~\cite{debenedetti2001supercooled}.
For quantitative analysis, we used the larger particles with $d=2.1$\,mm to conduct a series of experiments at various packing fractions, $\phi\in[0.88,0.91]$, using a relatively high frequency $f=65$\,Hz and low amplitude $A=0.052$\,mm, with respect to the parameter space covered in the phase diagram in Fig.~\ref{fig:phase}b.
The larger particles were chosen for the higher-frequency tests to approximate the limit when the supposed wavelength $\lambda\approx d$, which may suppress wave formation and wave-induced correlated particle motion. 
To obtain particle trajectories over time, we zoomed into a sub-region of the raft and captured images with a resolution of 864$\times$860\,pixels and a frame rate of 186\,Hz. A Gaussian filter was applied to the time series position data with a window size of five frames to construct particle trajectories.

For each packing fraction, we first calculated the particles' mean squared displacement (MSD) over time,
\[
\text{MSD}(\tau) = \langle |\mathbf{r}(t+\tau) - \mathbf{r}(t)|^2 \rangle_t,
\]
where $\mathbf{r}(t)$ is the position of a particle at a starting time $t$ and $\tau$ is the lag time over which the displacement is considered. In glassy systems, particles are sub-diffusive at intermediate $\tau$, reflecting the confinement of particles within cages. The relation between the MSD and $\tau$ can then be characterized as
\[
\text{MSD} \sim \tau^\alpha,
\]
with the exponent $\alpha \approx 0$ indicating a ``solid'' state, while $\alpha \approx 1$ indicates a ``liquid'' state with normal diffusion. 

\begin{figure}[h]
    \centering
    \includegraphics[width=0.975\linewidth]{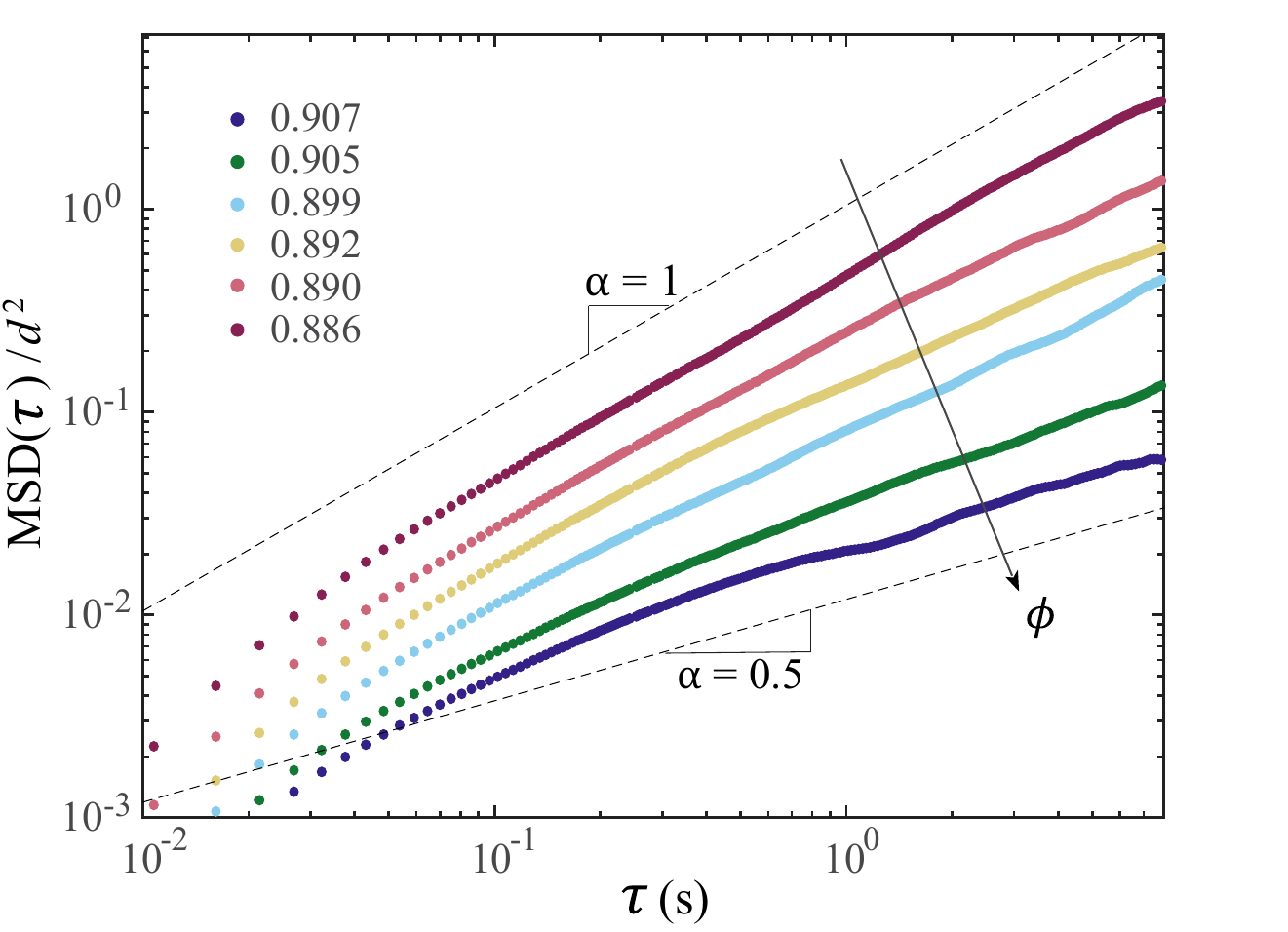}
    \caption{The mean squared displacement of $d=2.1$\,mm particles normalized by the squared particle diameter with various packing fractions at $f=65$\,Hz and $A=0.052$\,mm.}
    \label{fig:msdgraph}
\end{figure}

In the most dilute case considered, $\phi=0.886$, the particles have normal diffusive motion, which is made clear with the reference line of $\alpha = 1$ in the MSD plot in Fig.~\ref{fig:msdgraph}.
In the densest case considered, $\phi=0.907$, the behavior of the raft is clearly sub-diffusive, as $\alpha\approx0.5$, implying that the motion of individual particles is highly constrained by their neighbors within our duration of observation.
Thus, over a tight range of $\phi$, the vibrated raft exhibits qualitatively different behaviors, with the slowdown in its dynamics resembling that of a liquid-like particle system evolving toward its glass transition as it is compressed or cooled.

\begin{figure}[h]
    \centering
    \includegraphics[width=1.02\linewidth]{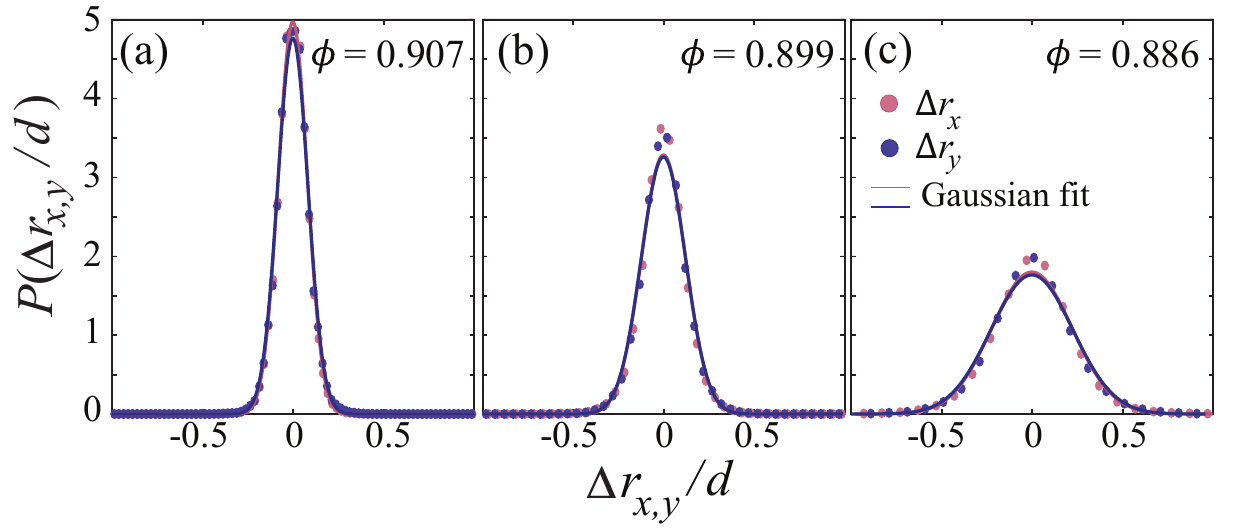}
    \caption{Probability distributions of the two displacement components over $\tau=0.33$\,s for $d=2.1$\,mm particles in a raft measured for $A=0.052$\,mm and $f=65$\,Hz at (a) $\phi=0.907$, (b) $\phi=0.899$, and (c) $\phi=0.886$. Bin sizes of (a) 0.025, (b) 0.05, and (c) 0.1 were used. A fitted Gaussian distribution is shown for each displacement component at each packing fraction.}
    \label{fig:velocitypdfs}
\end{figure}

We further consider the probability distribution of individual particle displacements, $\{\Delta r_x,\Delta r_y\}$, over a given lag time $\tau$.
We subtracted the mean from each distribution to account for the rigid body translation in the sub-region we monitored.
Figure~\ref{fig:velocitypdfs} shows the displacement distributions for three example packing fractions measured over $\tau=0.33$\,s, which can be well fitted to Gaussian distributions, indicating a resemblance to thermal particle motion.
This is likely a result of the diminishing wavelength of the standing waves and frequent particle collisions in this dense regime, leading to each particle experiencing rather random driving forces.

\begin{figure*}[t]
    \centering
    \includegraphics[width=1\linewidth]{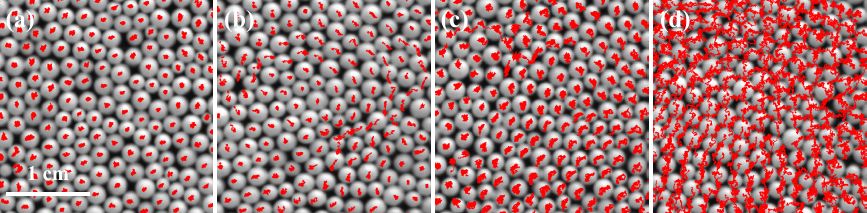}
    \caption{Particle trajectories for $d=2.1$\,mm particles over $\tau$=3\,s, measured for $A=0.052$\,mm and $f=65$\,Hz in (a) and (b) $\phi=0.907$, (c) $\phi=0.899$, and (d) $\phi=0.886$. 
    While certain regions in the $\phi=0.907$ case exhibit lower mobility, particles oscillating within effective cages imposed by surrounding particles as shown in (a), other regions in the same system over the same observation exhibit collective rearrangement, as shown in (b). For each case, see corresponding video in the supplementary~\cite{supp}.}
    \label{fig:densetrajectoriescollective}
\end{figure*}

Visualizations of individual particle trajectories over $\tau=3$\,s are shown in Fig.~\ref{fig:densetrajectoriescollective} for the three example packing fractions. 
At the highest packing fraction, Fig.~\ref{fig:densetrajectoriescollective}a, the particles mainly oscillate back and forth and rarely leave their effective cages. 
At the lowest packing fraction, all particles exhibit diffusive trajectories and have higher mobility within the same time period, as seen in Fig.~\ref{fig:densetrajectoriescollective}d. 
At the intermediate packing fraction, Fig.~\ref{fig:densetrajectoriescollective}c, while there is an observable effect from caging on particle trajectories, mobility is increased, and some particles have non-trivial displacements with correlation to those of their neighbors. This correlated motion is also present at the highest packing fraction, shown in Fig.~\ref{fig:densetrajectoriescollective}b, which shows a separate region of the system at the same time shown in Fig.~\ref{fig:densetrajectoriescollective}a, suggesting that the dynamics are spatially heterogeneous.

Dynamical heterogeneity is a hallmark of supercooled liquids close to the glass transition~\cite{ediger2000spatially} and confined granular particles under slow shear near the onset of jamming~\cite{dauchot2005dynamical,mehta2010spatial}, both of which are relevant to our current system. 
As such a system is compressed or experiences lower temperatures, its dynamics slow down and become more heterogeneous while simultaneously showing rapid growth of length- and time-scales in the correlation of particle motion.
This can be quantified by the four-point susceptibility, $\chi_4$, which assesses the correlation between the dynamics at any two spatial locations over a given time interval~\cite{keys2007measurement}. 
The calculation is based on measuring the overlap between two configurations.

For a system with $N$ particles, we first calculate the self-overlap order parameter over a delay time $\tau$, 
\[
q_{\mathrm{s}}(\tau)=\frac{1}{N}\sum_{i=1}^{N}w\left(|\mathbf{r}_i(\tau)-\mathbf{r}_i(0)|\right),
\]
where $\mathbf{r}_i$ is the particle position and $w$ is chosen to be a step function~\cite{keys2007measurement} such that $w=1$ if the particle displacement $|\mathbf{r}_i(\tau)-\mathbf{r}_i(0)|<0.3d$ and $w=0$ otherwise.
The susceptibility can then be calculated as
\[
\chi_4(\tau)=N\left(\left\langle q_\mathrm{s}^2\right\rangle-\left\langle q_\mathrm{s}\right\rangle^2\right),
\]
in which $\langle \cdot \rangle$ again denotes averaging over all starting times. Under this definition, a larger $\chi_4$ indicates stronger dynamical heterogeneity.  

\begin{figure}[b!]
    \centering
    \includegraphics[width=0.8\linewidth]{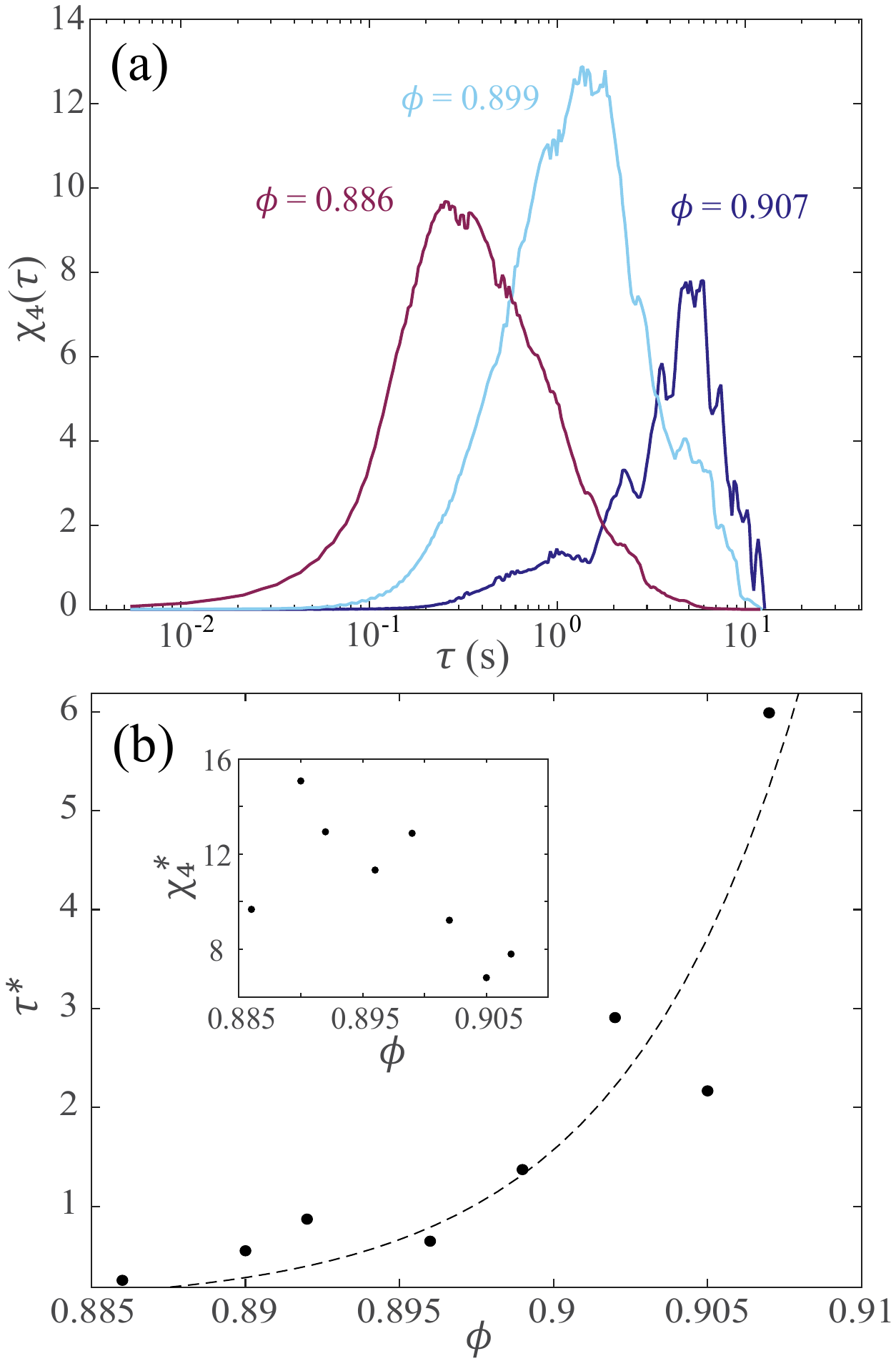}
    \caption{(a) The four-point susceptibility $\chi_4$ vs. delay time $\tau$ measured for three example packing fractions. (b) Dependence of the dynamical timescale $\tau^*$ on the packing fraction $\phi$. The inset in (b) shows the behavior of the peak value $\chi_4^*$ with increasing $\phi$.}
    \label{fig:dynamicaltimescale}
\end{figure}

The measured relation between $\chi_4$ and the delay time $\tau$ for a given packing fraction is similar to that of a glassy system, as shown in Fig.~\ref{fig:dynamicaltimescale}a. At small $\tau$, all particles vibrate within their respective cages, resulting in $\chi_4\approx0$, while at large $\tau$, all particles have diffused, also resulting in $\chi_4\approx0$.
At intermediate $\tau$, $\chi_4$ reaches a peak value at time $\tau^*$, setting a timescale for when dynamical heterogeneity is strongest. 
As the packing fraction increases, $\tau^*$ increases exponentially with $\phi$, see Fig.~\ref{fig:dynamicaltimescale}b, indicating rapid slowdown in the dynamics to develop correlated motion.

While the growth of $\tau^*$ fits the classical description of dynamical heterogeneity for glassy systems, the peak value of $\chi_4$ varies non-monotonically with $\phi$, as seen in Fig.~\ref{fig:dynamicaltimescale}a, which may reflect two unique features in our vibrated raft. First, as $\phi$ increases, the standing waves further diminish due to the more frequent particle collisions and the increased out-of-plane bending rigidity as in Fig.~\ref{fig:dispersion}d. The disappearance of the waves de-correlates the motion of nearby particles and thus decreases the dynamical heterogeneity. Second, at the vertically vibrated interface, a highly confined particle can move slightly out-of-plane, which may discourage the formation of strong contact force networks and thus de-correlate the motion of nearby particles.
This makes the effective interaction potential between particles softer and the raft more compressible. These two mechanisms may compete with the traditional role of density in the dynamical heterogeneity.

\section{Cavity formation in particle rafts under high-frequency and high-amplitude vibrations}
 
\begin{figure}[h!]
\centering
\includegraphics[width=0.99\linewidth]{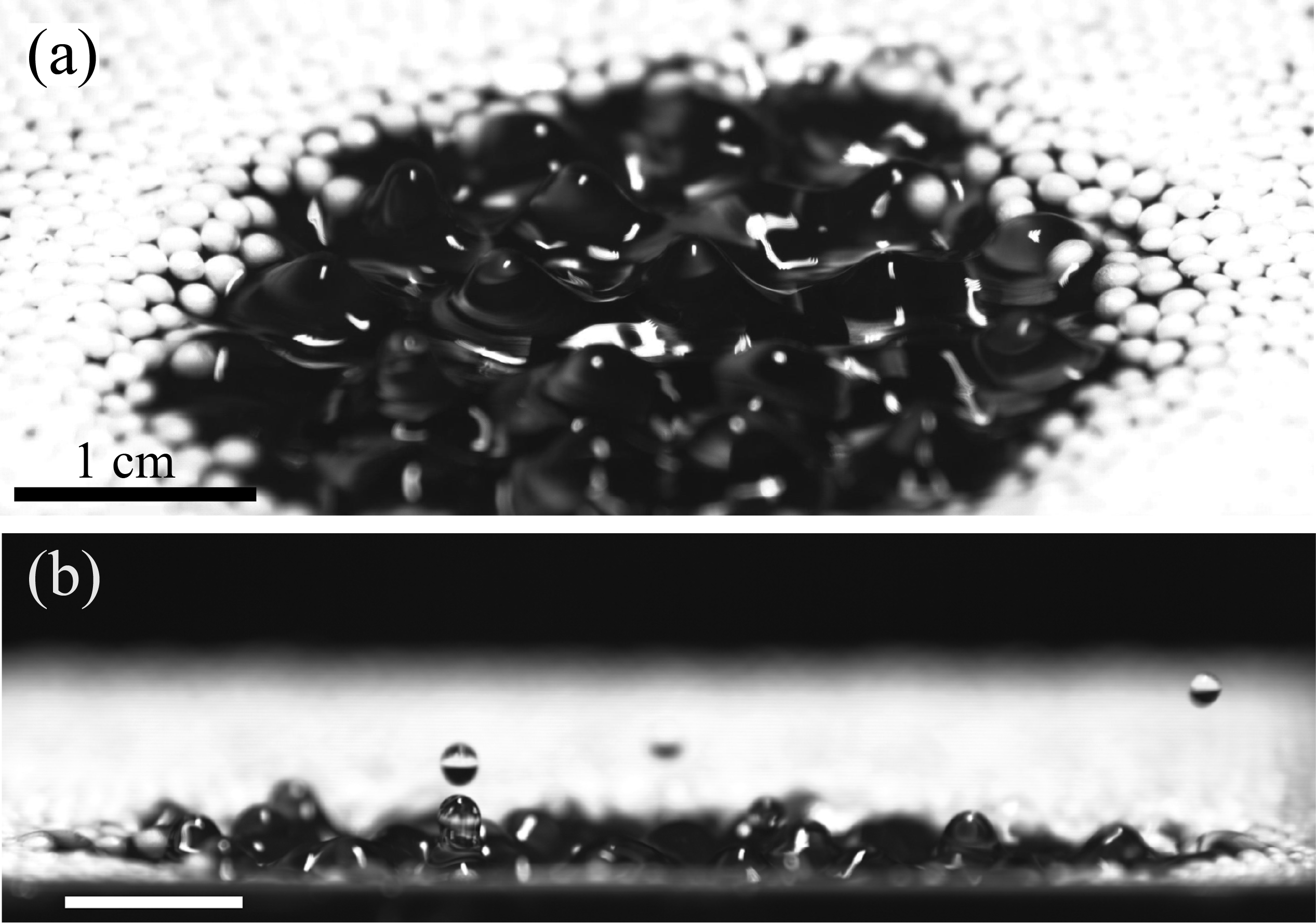}
\caption{Snapshots of the cavity and its surroundings, viewed from (a) an inclined perspective and (b) a side perspective. See corresponding videos in the supplementary~\cite{supp}.}
\label{fig:cavity}
\end{figure}

Starting from the glassy regime in the frequency range $f\in[75,95]$\,Hz, we reach the cavity-forming regime by increasing the vibration amplitude to a critical value $A^*$, at which a cavity nucleates and rapidly expands within a few seconds, before reaching a steady size and shape.  
The size of a cavity is represented by a diameter $L$ of an equivalent circle with the same area, which is at least one order of magnitude larger than the particle diameter, see high-speed images in Fig.~\ref{fig:cavity}a,b. Free-surface waves are observed inside the cavity, suggesting less damping there due to the lack of particles.

As the observed cavity is stable, the closing pressure from the surrounding dense raft should be balanced by a mechanism that drives particles outward. 
While individual particles occasionally re-enter the cavity, as seen in the upper right region of Fig.~\ref{fig:cavity}a, they quickly drift back to the surrounding dense region. 
Thus, the cavity's size is stable due to the balance between this drifting effect and the closing pressure of the surrounding raft.

The outward drifting motion of the particles is likely related to the particle-wave interaction. 
In our experiments, the Styrofoam particles tend to migrate away from liquid surfaces that experience intense vertical oscillations, i.e., the antinodes of the free-surface waves inside the cavity.
Due to this effect, a particle in the cavity may hop between the nodes of the free-surface waves and eventually end up ``trapped'' by the surrounding raft, which does not have standing waves.

Following this argument, a difference can be noticed between the cavitation regime and the regular wave regime. 
While particles also move away from the antinodes of the standing waves at lower frequencies in Fig.~\ref{fig:freqcomp} and Fig.~\ref{fig:phase}c-e, the wavelengths of those waves are so large that particles never evacuate a region spanning a full wavelength to allow a cavity with at least one free-surface wave to develop. As cavitation occurs at higher frequencies where the supposed wavelength is smaller, the physics may be different.

\begin{figure}[h!]
\centering
\includegraphics[width=\linewidth]{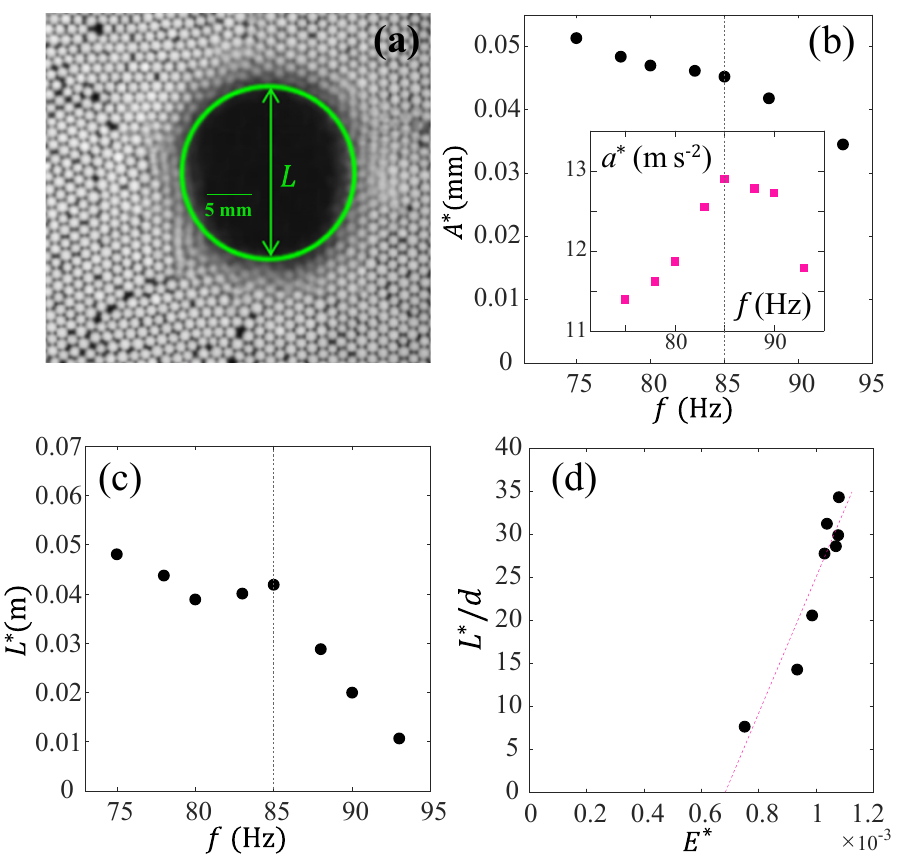}
\caption{Cavity formation for the $d=1.4$\,mm particles at packing fraction $\phi=0.90$. (a) A time-averaged experimental image for $d=1.4$\,mm, $\phi=0.90$, $f=90$\,Hz, and $A=0.029$\,mm, with the cavity identified. 
(b) Critical amplitude vs. frequency. The inset shows the critical acceleration vs. frequency.
(c) Critical cavity size vs. frequency. (d) Critical cavity size vs. the characteristic energy. The black circles are experimental measurements and the dashed line is a linear fit $L^*/d=7.9\times10^4E^*-54.1$. 
}
\label{fig:cavity_size}
\end{figure}

To understand the frequency dependence of the cavitation, we performed experiments using the smaller particles with $d=1.4$\,mm at various frequencies $f$.  
For each $f$, we gradually increased the vibration amplitude from the lower end until reaching the critical amplitude $A^*$, and the result in Fig.~\ref{fig:cavity_size}b shows a lower $A^*$ for higher $f$.

Given this, we speculate that the role of the amplitude in cavitation should be to induce large fluctuations in the local packing density so that a cavity with free-surface waves can nucleate.
For $A<A^*$, the sizes of the fluctuating voids in between particles in the dense packing are smaller than the wavelength of a free-surface wave, and those voids do not expand. When $A=A^*$, it is then possible that the size of a fluctuating void exceeds the wavelength of the free-surface wave, allowing the void to grow into a cavity that contains free-surface waves. 
As the wavelength is shorter for higher frequency, the required amplitude is smaller, hence the observation in Fig.~\ref{fig:cavity_size}b. 

As for the characteristics of the cavity, its stabilized size $L^*$ at the critical amplitude generally decreases with the vibration frequency, as shown in Fig.~\ref{fig:cavity_size}c. 
As the kinetic energy of the container is proportional to $A^2f^2$, we examine the relation between $L^*/d$ and a characteristic critical energy, $E^*=A^{*2}f^2/(gd)$. 
The result in Fig.~\ref{fig:cavity_size}d shows an approximately linear trend between $L^*/d$ and $E^*$, with a cut-off value around $E^*=7\times10^{-4}$, which may correspond to a high-frequency value at which surface waves always exist in the small voids dispersed in a disordered particle packing. 
Increasing the frequency beyond our range of interest will result in the formation of multiple cavities that are less well-defined and have smaller areas approximately equal to those of a few particles, while still containing surface waves or small ripples.
As noted earlier, further increasing the amplitude beyond $A^*$ will result in the cavity expanding further, and water droplets may be ejected (similar to Fig.~\ref{fig:cavity}b).

Lastly, we point out that a small increasing trend in the critical cavity size $L^*$ exists as $f$ approaches 85\,Hz, which can be seen in Fig.~\ref{fig:cavity_size}c. This specific frequency coincides with the frequency at which the critical vibration acceleration, $a^*=4\pi^2f^2A^*$, is largest, see the inset of Fig.~\ref{fig:cavity_size}b. This non-monotonic behavior likely indicates a change in the physics of the raft around the cavity. 
For $f<85$\,Hz, the cavity assumes an irregular shape that changes slightly over time, especially at smaller packing fractions.
In contrast, for $f\ge85$\,Hz, the cavity maintains a circular and stable shape, as in Fig.~\ref{fig:cavity_size}a for $f=90$\,Hz. 
In particular, Fig.~\ref{fig:cavity_size}a is a time-averaged image, showing that particles at the perimeter of the cavity have large fluctuations in their positions as the image is blurry. Beyond the cavity's perimeter, particles that are driven outward can stack upon one another, forming a ``brighter ring'' in the image, corresponding to a 2D local packing fraction around 0.95, measured from the stacked particles' projection.
For particles located approximately ten layers further away from the cavity, the corresponding image is not blurry, as particles and the dispersed small voids appear distinctly, indicating a strong caging effect. This means the circular-shaped cavitation occurs against elastic-like resistance from the raft in a glassy state, whereas the irregular-shaped cavity expands under little resistance from the raft in a liquid-like state.

\section{Conclusion and outlook}

Our work demonstrates that particle packing fraction, vibration frequency, and amplitude serve as important control parameters for the dynamic behavior of vertically vibrated granular rafts. By systematically tuning these parameters, we map out a phase diagram of five regimes spanning from classical Faraday wave-dominated behaviors to particle dynamics-dominated behaviors such as glassy particle dynamics and cavitation.
Their occurrence and transitions are governed by the interplay of the hydrodynamics of vibrated liquid interfaces, capillary effects, and the dissipative particle interactions, which at higher packing densities involve elastic-like mechanical response and glassy particle dynamics. 

Beyond the parameters we varied, we expect the surface tension, contact angle, and particle and liquid densities to also be important for the behavior of a dense granular raft under vibration.
These factors, along with particle size, determine the vertical position of particles with respect to the interface and how nearby particles collectively distort the liquid-air interface~\cite{dalbe2011aggregation}; we note that the surface tension and liquid density, along with several other parameters (e.g., liquid depth and gravitational acceleration), determine the dynamics of a vibrated pure liquid system~\cite{lioubashevski1997scaling}.
In our current setting, the contact angle for the Styrofoam particles lies at the threshold between wetting and non-wetting, the density of the particles is much smaller than that of water, and the particle size is comparable to the capillary length. 
Future studies could manipulate these parameters to probe the roles of the respective effects in the vibration-induced dynamics and in the raft properties.
For example, with decreased contact angle and increased particle density,  particles may be more submerged in the liquid and induce larger distortions to the interface, possibly resulting in more damped particle dynamics, more added inertia to the interface, and increased bending rigidity of the raft.
Understanding these effects will have important implications for the creation of dynamically responsive and configurable particle systems. 

Apart from the interfacial fluid-mechanical aspect, the vertically vibrated raft can be a model particle system for understanding out-of-equilibrium physics, such as the glass transition, the deformation of amorphous solids at finite (effective) temperatures, and the dynamics of polycrystals if monodisperse particles are used. Different from the commonly considered interaction potentials such as the Lennard-Jones and hard-sphere potentials, the interactions between particles at vibrated interfaces are complex, with a many-body nature that arises from the capillary and hydrodynamic effects. As mentioned above, such interactions can be controlled by manipulating the particle and liquid properties~\cite{dalbe2011aggregation}, and additional wave-driven particle dynamics can be introduced via manipulating particle shapes~\cite{Barotta2023,Sungar2025,Barotta2025}.
Being able to experimentally control these complex particle-scale features and introduce dynamical excitations can be important in extending our fundamental understanding of the dynamics and properties of dense particle assemblies from idealized to real-world systems.

\begin{acknowledgments}
The authors would like to acknowledge the funding from the National Science Foundation grant CMMI-2519512. We thank Yue Fan, Richard Lueptow, and Paul Umbanhowar for helpful discussions.
\end{acknowledgments}

\bibliography{apssamp}

@article{nicolson1949interaction,
   title={The Interaction Between Floating Particles},
  author={Nicolson, MM},
  journal = {Proc. Cambridge Philos. Soc.},
  volume={45},
  number={2},
  pages={288-295},
  year={1949},
  DOI = {10.1017/S0305004100024841}
}

@article{kralchevsky2000capillary,
  title={Capillary interactions between particles bound to interfaces, liquid films and biomembranes},
  author={Kralchevsky, PA and Nagayama, K},
  journal={Adv. Colloid Interface Sci.},
  volume={85},
  number={2-3},
  pages={145-192},
  year={2000},
  DOI = {10.1016/S0001-8686(99)00016-0},
  ISSN = {0001-8686},
  EISSN = {1873-3727},
  ResearcherID-Numbers = {Nagayama, Kuniaki/O-2522-2018
   Kralchevsky, Peter/A-1012-2008},
  ORCID-Numbers = {Kralchevsky, Peter/0000-0003-3942-1411}
}

@article{vella2005cheerios,
  title={The “cheerios effect”},
  author={Vella, D and Mahadevan, L},
  journal={Am. J. Phys.},
  volume={73},
  number={9},
  pages={817-825},
  year={2005},
  Month = {SEP},
  DOI = {10.1119/1.1898523},
  ISSN = {0002-9505},
  EISSN = {1943-2909},
  ResearcherID-Numbers = {Vella, Dominic/A-9070-2008
   },
ORCID-Numbers = {Vella, Dominic/0000-0003-1341-8863
   Mahadevan, Lakshminarayanan/0000-0002-5114-0519}
}

@misc{supp,
  note = {See the supplementary videos at [URL will be inserted by publisher].},
}

@article{Barotta2023,
  author = {Barotta, Jack-William and Thomson, Stuart J. and Alventosa, Luke F. L.
   and Lewis, Maya and Harris, Daniel M.},
  title = {Bidirectional Wave-Propelled Capillary Spinners},
  journal = {Commun. Phys.},
  volume = {6},
  number = {1},
  pages={87},
  year = {2023},
  DOI = {10.1038/s42005-023-01206-z},
  ISSN = {2399-3650},
  ORCID-Numbers = {Barotta, Jack-William/0000-0001-6769-5132
   Thoon, Stuart/0000-0003-1101-6457}
}

@article{Sungar2025,
  author = {Sungar, Nilgun and Sharpe, John and Ijzerman, Loic and Barotta,
   Jack-William},
  title = {Synchronization and Self-Assembly of Free Capillary Spinners},
  journal = {Phys. Rev. E},
  volume = {111},
  number = {3},
  pages={035104},
  year = {2025},
  DOI = {10.1103/PhysRevE.111.035104},
  ISSN = {2470-0045},
  EISSN = {2470-0053},
  ORCID-Numbers = {Barotta, Jack-William/0000-0001-6769-5132
   Sungar, Nilgun/0009-0003-7270-3602
   Ijzerman, Loic/0009-0007-7866-6805}
}

@article{Barotta2025,
  author = {Barotta, Jack-William and Pucci, Giuseppe and Silver, Eli and
   Hooshanginejad, Alireza and Harris, Daniel M.},
  title = {Synchronization of Wave-Propelled Capillary Spinners},
  journal = {Phys. Rev. E},
  volume = {111},
  number = {3},
  pages = {035105},
  year = {2025},
  DOI = {10.1103/PhysRevE.111.035105},
  ISSN = {2470-0045},
  EISSN = {2470-0053},
  ResearcherID-Numbers = {Pucci, Giuseppe/W-2613-2018},
  ORCID-Numbers = {Barotta, Jack-William/0000-0001-6769-5132
   Pucci, Giuseppe/0000-0001-6621-3277
   Silver, Eli/0009-0005-9984-5760}
}

@article{to2023rifts,
  title={Rifts in rafts},
  author={T{\^o}, Kh{\'a}-{\^I} and Nagel, Sidney R},
  journal={Soft Matter},
  volume={19},
  number={5},
  pages={905-912},
  year={2023},
  Month = {FEB 1},
  DOI = {10.1039/d2sm01451c},
  EarlyAccessDate = {JAN 2023},
  ISSN = {1744-683X},
  EISSN = {1744-6848},
  ORCID-Numbers = {To, Kha-I/0000-0001-7602-7822},
}

@article{dalbe2011aggregation,
  title={Aggregation of frictional particles due to capillary attraction},
  author={Dalbe, Marie-Julie and Cosic, Darija and Berhanu, Michael and Kudrolli, Arshad},
  journal={Phys. Rev. E},
  volume={83},
  number={5},
  year={2011},
  Month = {MAY 23},
  DOI = {10.1103/PhysRevE.83.051403},
  pages = {051403},
  ISSN = {1539-3755},
  EISSN = {1550-2376},
  ResearcherID-Numbers = {Berhanu, Michael/AAV-9303-2020
   Kudrolli, Arshad/ABD-2656-2020
   },
  ORCID-Numbers = {Berhanu, Michael/0000-0001-9099-2135
   Kuolli, Arshad/0000-0003-1754-1519
   Dalbe, Marie-Julie/0000-0003-4300-8152},
}

@article{dani2015hydrodynamics,
  title={Hydrodynamics of Particles at an Oil-Water Interface},
  author={Dani, Archit and Keiser, Geoff and Yeganeh, Mohsen and Maldarelli, Charles},
  journal={Langmuir},
  volume={31},
  number={49},
  pages={13290-13302},
  year={2015},
  Month = {DEC 15},
  DOI = {10.1021/acs.langmuir.5b02146},
  ISSN = {0743-7463},
  ResearcherID-Numbers = {Dani, Archit/AET-3555-2022}
}

@article{protiere2017sinking,
  title={Sinking a granular raft},
  author={Proti{\`e}re, Suzie and Josserand, Christophe and Aristoff, Jeffrey M. and
   Stone, Howard A. and Abkarian, Manouk},
  journal={Phys. Rev. Lett.},
  volume={118},
  number={10},
  year={2017},
  doi = {10.1103/PhysRevLett.118.108001},
  pages = {108001},
  ISSN = {0031-9007},
  EISSN = {1079-7114},
  ResearcherID-Numbers = {Stone, Howard/B-6435-2008
   Abkarian, Manouk/OIS-5108-2025},
  ORCID-Numbers = {ABKARIAN, Manouk/0000-0003-0411-3187
   },
  publisher={APS}
}

@article{abkarian2013gravity,
  title={Gravity-induced encapsulation of liquids by destabilization of granular rafts},
  author={Abkarian, Manouk and Proti{\`e}re, Suzie and Aristoff, Jeffrey M and Stone, Howard A.},
  journal={Nat. Commun.},
  volume={4},
  number={1},
  pages={1895},
  year={2013},
  publisher={Nature Publishing Group UK London},
  DOI = {10.1038/ncomms2869},
  ISSN = {2041-1723},
  ResearcherID-Numbers = {Abkarian, Manouk/OIS-5108-2025
   Stone, Howard/B-6435-2008
   },
  ORCID-Numbers = {ABKARIAN, Manouk/0000-0003-0411-3187}
}

@article{protiere2023particle,
  title={Particle Rafts and Armored Droplets},
  author={Proti{\`e}re, Suzie},
  journal={Annu. Rev. Fluid Mech.},
  volume={55},
  pages={459-480},
  DOI = {10.1146/annurev-fluid-030322-015150},
  ISSN = {0066-4189},
  EISSN = {1545-4479},
  year={2023},
  publisher={Annual Reviews}
}

@article{druecke2023collapse,
  title={Collapse of a granular raft: Transition from single particle falling to collective creasing},
  author={Druecke, Benjamin C. and Mukherjee, Ranit and Cheng, Xiang and Lee,
   Sungyon},
  journal={Phys. Rev. Fluids},
  volume={8},
  number={2},
  Month = {FEB 22},
  DOI = {10.1103/PhysRevFluids.8.024003},
  pages = {024003},
  ISSN = {2469-990X},
  ResearcherID-Numbers = {Cheng, Xiang/G-5704-2017
   Mukherjee, Ranit/AAE-8494-2021},
  ORCID-Numbers = {Lee, Sungyon/0000-0002-4118-1712
   Cheng, Xiang/0000-0002-2759-764X
   },
  year={2023},
  publisher={APS}
}

@article{sayyari2025destabilizing,
  title={Destabilizing a buoyant multilayer granular raft by heavy grains: The role of inertia},
  author={Sayyari, Mohammad Javad and Bostwick, Joshua B.},
  journal={Langmuir},
  volume={41},
  number={28},
  pages={18403--18413},
  year={2025},
  publisher={ACS Publications},
  Month = {JUL 22},
  DOI = {10.1021/acs.langmuir.5c00012},
  EarlyAccessDate = {JUL 2025},
  ISSN = {0743-7463},
  EISSN = {1520-5827}
}

@article{douady1990experimental,
  title={Experimental study of the Faraday instability},
  author={Douady, S.},
  journal={J. Fluid Mech.},
  volume={221},
  pages={383-409},
  year={1990},
  DOI = {10.1017/S0022112090003603},
  ISSN = {0022-1120},
  ResearcherID-Numbers = {DOUADY, Stephane/C-4498-2018}
}

@article{Pak1993,
  author = {Pak, HK and Behringer, RP},
  title = {Surface waves in vertically vibrated granular materials},
  journal = {Phys. Rev. Lett.},
  volume = {71},
  number = {12},
  pages = {1832-1835},
  year = {1993},
  DOI = {10.1103/PhysRevLett.71.1832},
  ISSN = {0031-9007},
  ResearcherID-Numbers = {Pak, Hyuk/H-3727-2012}
}

@article{Lagarde2020, 
   author = {Lagarde, Antoine and Proti{\`e}re, Suzie}, 
   title = {Probing the erosion and cohesion of a granular raft in motion}, 
   journal = {Phys. Rev. Fluids}, 
   volume = {5}, 
   number = {4}, 
   pages = {044003}, 
   year = {2020}, 
   DOI = {10.1103/PhysRevFluids.5.044003},
   ISSN = {2469-990X},
   ORCID-Numbers = {Lagarde, Antoine/0000-0001-8053-5803}
}

@article{edwards1994patterns,
  title={Patterns and quasi-patterns in the Faraday experiment},
  author={Edwards, W. S. and Fauve, S.},
  journal={J. Fluid Mech.},
  volume={278},
  pages={123-148},
  year={1994},
  DOI = {10.1017/S0022112094003642},
  ISSN = {0022-1120},
  EISSN = {1469-7645}
}

@article{batson2013parametric,
  Author = {Batson, W. and Zoueshtiagh, F. and Narayanan, R.},
  Title = {The Faraday threshold in small cylinders and the sidewall non-ideality},
  Journal = {J. Fluid Mech.},
  Year = {2013},
  Volume = {729},
  Pages = {496-523},
  Month = {AUG},
  DOI = {10.1017/jfm.2013.324},
  ISSN = {0022-1120},
  EISSN = {1469-7645},
  ORCID-Numbers = {Narayanan, Ranga/0000-0003-2394-5077
   Zoueshtiagh, Farzam/0000-0001-7642-6356
   Batson, William/0000-0002-2156-0272}
}

@article{faraday1830peculiar,
  title={On a peculiar class of acoustical figures; and on certain forms assumed by groups of particles upon vibrating elastic surfaces},
  author={Faraday, Michael},
  journal={Philos. Trans. R. Soc. London},
  volume={121},
  pages={299-340},
  year={1831},
  doi = {10.1098/rstl.1831.0018},
  publisher={The Royal Society London}
}

@article{miles1993faraday,
  title={On faraday waves},
  author={Miles, John},
  journal={J. Fluid Mech.},
  volume={248},
  pages={671-683},
  year={1993},
  DOI = {10.1017/S0022112093000965},
  ISSN = {0022-1120},
  publisher={Cambridge University Press}
}

@article{kumar1994parametric,
  title={Parametric instability of the interface between two fluids},
  author={Kumar, K. and Tuckerman, L. S.},
  journal={J. Fluid Mech.},
  volume={279},
  pages={49-68},
  year={1994},
  DOI = {10.1017/S0022112094003812},
  ISSN = {0022-1120}
}

@article{benjamin1954instability,
  title={The stability of the plane free surface of a liquid in vertical periodic motion},
  author={Benjamin, T. B. and Ursell, F.},
  journal={Proc. R. Soc. A},
  volume={225},
  pages={505-515},
  year={1954},
  DOI={10.1098/rspa.1954.0218}
}

@article{planchette2012surface,
  title={Surface wave on a particle raft},
  author={Planchette, Carole and Lorenceau, Elise and Biance, Anne-Laure},
  journal={Soft Matter},
  volume={8},
  number={8},
  pages={2444-2451},
  year={2012},
  publisher={Royal Society of Chemistry},
  DOI = {10.1039/c2sm06859a},
  ISSN = {1744-683X},
  ResearcherID-Numbers = {Planchette, Carole/G-6266-2018
   Biance, Anne-Laure/J-8884-2017},
  ORCID-Numbers = {Biance, Anne-Laure/0000-0002-3120-7595
   Planchette, Carole/0000-0002-3974-5742
   }
}

@article{xiao2023identifying,
  title={Identifying microscopic factors that influence ductility in disordered solids},
  author={Xiao, Hongyi and Zhang, Ge and Yang, Entao and Ivancic, Robert and Ridout, Sean and Riggleman, Robert and Durian, Douglas J and Liu, Andrea J},
  journal={Proc. Natl. Acad. Sci. U. S. A.},
  volume={120},
  number={42},
  pages={e2307552120},
  year={2023},
  DOI = {10.1073/pnas.2307552120},
  ISSN = {0027-8424},
  EISSN = {1091-6490},
  ResearcherID-Numbers = {Liu, Andrea/K-3968-2017
   xiao, hongyi/GPX-7219-2022
   Liu, Anea/K-3968-2017
   Zhang, Ge/AAQ-2032-2020
   Durian, Douglas/D-7013-2012
   Ivancic, Robert Joseph/AAN-8990-2021
   Yang, Entao/JHU-0878-2023
   },
  ORCID-Numbers = {Liu, Anea/0000-0002-2295-2729
   Xiao, Hongyi/0000-0001-5174-2061
   Yang, Entao/0000-0002-0420-8920
   Durian, Douglas/0000-0003-3240-2381
   Zhang, Ge/0000-0002-6377-7760
   Ridout, Sean/0000-0003-2387-8361},
   publisher={National Academy of Sciences}
}

@article{xiao2020strain,
  title={Strain localization and failure of disordered particle rafts with tunable ductility during tensile deformation},
  author={Xiao, Hongyi and Ivancic, Robert JS and Durian, Douglas J},
  journal={Soft Matter},
  volume={16},
  number={35},
  pages={8226-8236},
  year={2020},
  DOI = {10.1039/d0sm00839g},
  ISSN = {1744-683X},
  EISSN = {1744-6848},
  ResearcherID-Numbers = {Ivancic, Robert Joseph/AAN-8990-2021
   Xiao, Hongyi/AAM-5333-2020
   Durian, Douglas/D-7013-2012},
  ORCID-Numbers = {Xiao, Hongyi/0000-0001-5174-2061
   Ivancic, Robert/0000-0001-9969-2534
   Durian, Douglas/0000-0003-3240-2381},
  publisher={Royal Society of Chemistry}
}

@article{atherton1999detection,
  title={Size invariant circle detection},
  author={Atherton, T.J. and Kerbyson, D.J.},
  journal={Image Vis. Comput.},
  volume={17},
  number={11},
  pages={795-803},
  year={1999},
  DOI={10.1016/S0262-8856(98)00160-7},
  ISSN = {0262-8856},
}

@article{saddier2024breaking,
  title={Breaking of a floating particle raft by water waves},
  author={Saddier, Louis and Palotai, Ambre and Aksil, Math{\'e}o and Tsamados, Michel and Berhanu, Michael},
  journal={Phys. Rev. Fluids},
  volume={9},
  number={9},
  pages={094302},
  year={2024},
  publisher={APS},
  DOI={10.1103/PhysRevFluids.9.094302}
}

@article{umbanhowar1996localized,
  title={Localized excitations in a vertically vibrated granular layer},
  author={Umbanhowar, Paul B and Melo, Francisco and Swinney, Harry L},
  journal={Nature},
  volume={382},
  number={6594},
  pages={793--796},
  year={1996},
  publisher={Nature Publishing Group UK London},
  DOI={10.1038/382793a0}
}

@article{melo1994transition,
  title={Transition to parametric wave patterns in a vertically oscillated granular layer},
  author={Melo, Francisco and Umbanhowar, Paul and Swinney, Harry L},
  journal={Phys. Rev. Lett.},
  volume={72},
  number={1},
  pages={172},
  year={1994},
  publisher={APS},
  DOI={10.1103/PhysRevLett.72.172}
}

@book{landau2012theory,
  title={Theory of elasticity: volume 7},
  author={Landau, Lev Davidovich and Pitaevskii, LP and Kosevich, Arnold Markovich and Lifshitz, Evgenii Mikhailovich},
  volume={7},
  year={2012},
  publisher={Elsevier}
}

@article{debenedetti2001supercooled,
  title={Supercooled liquids and the glass transition},
  author={Debenedetti, Pablo G and Stillinger, Frank H},
  journal={Nature},
  volume={410},
  number={6825},
  pages={259--267},
  year={2001},
  publisher={Nature Publishing Group UK London},
  DOI={10.1038/35065704}
}

@article{kim2019failure,
  author = {Kim, Brian L. and Rendos, Abigail and Ganesh, Prithika and Brown, Keith
   A.},
  title = {Failure of Particle-Laden Interfaces Studied Using The Funnel Method},
  journal = {Colloid Interface Sci. Commun.},
  year = {2019},
  volume = {28}, 
  pages = {54-59},
  month = {JAN},
  DOI = {10.1016/j.colcom.2018.11.008},
  ISSN = {2215-0382},
  ResearcherID-Numbers = {Brown, Keith/A-9780-2010},
  ORCID-Numbers = {Kim, Brian L/0000-0002-2403-8703
   Brown, Keith/0000-0002-2379-2018}
}

@article{Vella2004212,
  author = {Vella, D. and Aussillous, P. and Mahadevan, L.},
  title = {Elasticity of an interfacial particle raft},
  year = {2004},
  journal = {Europhys. Lett.},
  volume = {68},
  number = {2},
  pages = {212 – 218},
  DOI = {10.1209/epl/i2004-10202-x},
  url = {https://www.scopus.com/inward/record.uri?eid=2-s2.0-7444247993&doi=10.1209%2fepl%2fi2004-10202-x&partnerID=40&md5=de631d9c3e84f6260a851c84b0d3ba97}
}

@article{Aveyard20001969,
  author = {Aveyard, Robert and Clint, John H. and Nees, Dieter and Paunov, Vesselin N.},
  title = {Compression and structure of monolayers of charged latex particles at air/water and octane/water interfaces},
  year = {2000},
  journal = {Langmuir},
  volume = {16},
  number = {4},
  pages = {1969 – 1979},
  DOI = {10.1021/la990887g},
  url = {https://www.scopus.com/inward/record.uri?eid=2-s2.0-0033897994&doi=10.1021%2fla990887g&partnerID=40&md5=dbdf5a419925bb1183fdeddf4dd4a0f6}
}

@article{vella2006dynamics,
  title = {Dynamics of Surfactant-Driven Fracture of Particle Rafts},
  author = {Vella, Dominic and Kim, Ho-Young and Aussillous, Pascale and Mahadevan, L.},
  journal = {Phys. Rev. Lett.},
  volume = {96},
  issue = {17},
  pages = {178301},
  numpages = {4},
  year = {2006},
  month = {May},
  publisher = {American Physical Society},
  doi = {10.1103/PhysRevLett.96.178301},
  url = {https://link.aps.org/doi/10.1103/PhysRevLett.96.178301}
}

@article{aveyard2000structure,
  author = {Aveyard, Robert and Clint, John H. and Nees, Dieter and Quirke, Nick},
  title = {Structure and Collapse of Particle Monolayers under Lateral Pressure at the Octane/Aqueous Surfactant Solution Interface},
  journal = {Langmuir},
  volume = {16},
  number = {23},
  pages = {8820-8828},
  year = {2000},
  doi = {10.1021/la000060i},
  URL = {https://doi.org/10.1021/la000060i}
}

@article{zang2010viscoelastic,
  author = {Zang, D. Y. and Rio, E. and Langevin, D. and Wei, B. and Binks, B. P.},
  title = {Viscoelastic properties of silica nanoparticle monolayers at the
   air-water interface},
  journal = {Eur. Phys. J. E: Soft Matter Biol. Phys.},
  year = {2010},
  volume = {31},
  number = {2},
  pages = {125-134},
  month = {FEB},
  DOI = {10.1140/epje/i2010-10565-7},
  ISSN = {1292-8941},
  EISSN = {1292-895X},
  ResearcherID-Numbers = {Binks, Bernard/JYP-5516-2024},
  ORCID-Numbers = {Zang, Duyang/0000-0003-4547-0957}
}

@article{kralchevsky2005,
  author = {Kralchevsky, PA and Ivanov, IB and Ananthapadmanabhan, KP and Lips, A},
  title = {On the thermodynamics of particle-stabilized emulsions: Curvature
   effects and catastrophic phase inversion},
  journal = {Langmuir},
  year = {2005},
  volume = {21},
  pages = {50-63},
  month = {JAN 4},
  DOI = {10.1021/la047793d},
  ISSN = {0743-7463},
  EISSN = {1520-5827},
  ResearcherID-Numbers = {Kralchevsky, Peter/A-1012-2008
   Lips, Alex/AAF-3408-2020
   Ivanov, Ivan/J-3894-2013},
  ORCID-Numbers = {Lips, Alexander/0000-0002-7538-5430
   Kralchevsky, Peter/0000-0003-3942-1411
   }
}

@article{aveyard2003,
  author = {Aveyard, R and Clint, JH and Horozov, TS},
  title = {Aspects of the stabilisation of emulsions by solid particles: Effects of
   line tension and monolayer curvature energy},
  journal = {Phys. Chem. Chem. Phys.},
  year = {2003},
  volume = {5},
  pages = {2398-2409},
  DOI = {10.1039/b210687f},
  ISSN = {1463-9076},
  EISSN = {1463-9084},
  ResearcherID-Numbers = {Horozov, Tommy/C-6417-2008},
  ORCID-Numbers = {Horozov, Tommy/0000-0001-8818-3750}
}

@article{danov2001capillary,
  author = {Danov, KD and Pouligny, B and Kralchevsky, PA},
  title = {Capillary forces between colloidal particles confined in a liquid film:
   The finite-meniscus problem},
  journal = {Langmuir},
  year = {2001},
  volume = {17},
  pages = {6599-6609},
  month = {OCT 16},
  DOI = {10.1021/la0107300},
  ISSN = {0743-7463},
  ResearcherID-Numbers = {Kralchevsky, Peter/A-1012-2008
   Danov, Krasimir/C-2220-2018},
  ORCID-Numbers = {Kralchevsky, Peter/0000-0003-3942-1411
   Danov, Krasimir/0000-0002-9563-0974}
}

@article{peco2017surfacetension,
  author = {Peco, Christian and Chen, Wei and Liu, Yingjie and Bandi, M. M. and
   Dolbow, John E. and Fried, Eliot},
  title = {Influence of surface tension in the surfactant-driven fracture of
   closely-packed particulate monolayers},
  journal = {Soft Matter},
  year = {2017},
  volume = {13},
  pages = {5832-5841},
  month = {SEP 21},
  DOI = {10.1039/c7sm01245d},
  ISSN = {1744-683X},
  EISSN = {1744-6848},
  ResearcherID-Numbers = {Liu, Yingjie/F-1888-2015
   Fried, Eliot/B-5836-2015
   Dolbow, John/F-9347-2010},
  ORCID-Numbers = {Liu, Yingjie/0000-0002-1169-6213
   Dolbow, John/0000-0002-6250-2183
   Bandi, Mahesh/0000-0002-8431-9005
   Fried, Eliot/0000-0001-5329-3394
   }
}

@article{zhang2023pattern,
  author = {Zhang, Shimin and Borthwick, Alistair G. L. and Lin, Zhiliang},
  title = {Pattern evolution and modal decomposition of Faraday waves in a brimful
   cylinder},
  journal = {J. Fluid Mech.},
  year = {2023},
  volume = {974},
  pages={A56},
  DOI = {10.1017/jfm.2023.838},
  ISSN = {0022-1120},
  EISSN = {1469-7645},
  ORCID-Numbers = {Zhang, Shimin/0000-0002-3984-2358}
}

@article{shao2021surface,
  author = {Shao, X. and Wilson, P. and Saylor, J. R. and Bostwick, J. B.},
  title = {Surface wave pattern formation in a cylindrical container},
  journal = {J. Fluid Mech.},
  year = {2021},
  volume = {915},
  pages={A19},
  DOI = {10.1017/jfm.2021.97},
  ISSN = {0022-1120},
  EISSN = {1469-7645}
}

@article{ap2006properties,
  author = {Aussillous, P and Quéré, D},
  title = {Properties of liquid marbles},
  journal = {Proc. R. Soc. London A},
  year = {2006},
  volume = {462},
  pages = {973-999},
  DOI = {10.1098/rspa.2005.1581},
  ISSN = {1364-5021},
  ResearcherID-Numbers = {Aussillous, Pascale/HJP-4762-2023
   Quere, David/KLD-5230-2024}
}

@article{dauchot2005dynamical,
  title={Dynamical heterogeneity close to the jamming transition in a sheared granular material},
  author={Dauchot, Olivier and Marty, Guillaume and Biroli, Giulio},
  journal={Phys. Rev. Lett.},
  volume={95},
  number={26},
  pages={265701},
  year={2005},
  DOI = {10.1103/PhysRevLett.95.265701},
  publisher={APS}
}

@article{keys2007measurement,
  title={Measurement of growing dynamical length scales and prediction of the jamming transition in a granular material},
  author={Keys, Aaron S and Abate, Adam R and Glotzer, Sharon C and Durian, Douglas J},
  journal={Nat. Phys.},
  volume={3},
  number={4},
  pages={260--264},
  year={2007},
  DOI = {10.1038/nphys572},
  publisher={Nature Publishing Group UK London}
}

@article{mlot2011fire,
  title={Fire ants self-assemble into waterproof rafts to survive floods},
  author={Mlot, Nathan J and Tovey, Craig A and Hu, David L},
  journal={Proc. Natl. Acad. Sci. U. S. A.},
  volume={108},
  number={19},
  pages={7669--7673},
  year={2011},
  DOI = {10.1073/pnas.1016658108},
  publisher={National Academy of Sciences}
}

@article{hostache1998reproductive,
  title={Reproductive biology of the neotropical armoured catfish Hoplosternum littorale (Siluriformes—Callichthyidae): a synthesis stressing the role of the floating bubble nest},
  author={Hostache, G{\'e}rard and Mol, Jan H},
  journal={Aquat. Living Resour.},
  volume={11},
  number={3},
  pages={173--185},
  year={1998},
  DOI = {10.1016/S0990-7440(98)80114-9},
  publisher={Elsevier}
}

@article{tan2022odd,
  title={Odd dynamics of living chiral crystals},
  author={Tan, Tzer Han and Mietke, Alexander and Li, Junang and Chen, Yuchao and Higinbotham, Hugh and Foster, Peter J and Gokhale, Shreyas and Dunkel, J{\"o}rn and Fakhri, Nikta},
  journal={Nature},
  volume={607},
  number={7918},
  pages={287--293},
  year={2022},
  DOI = {10.1038/s41586-022-04889-6},
  publisher={Nature Publishing Group UK London}
}

@article{lioubashevski1997scaling,
  title={Scaling of the transition to parametrically driven surface waves in highly dissipative systems},
  author={Lioubashevski, O and Fineberg, J and Tuckerman, LS},
  journal={Phys. Rev. E},
  volume={55},
  number={4},
  pages={R3832},
  year={1997},
  publisher={APS},
  DOI = {10.1103/PhysRevE.55.R3832}
}

@article{ezersii1986spatiotemporal,
  title={Spatiotemporal chaos in the parametric excitation of a capillary ripple},
  author={Ezerskii, AB and Rabinovich, MI and Reutov, VP and Starobinets, IM},
  journal={JETP},
  volume={64},
  number={6},
  pages={1228},
  year={1986}
}

@article{shani2010localized,
  title={Localized instability on the route to disorder in Faraday waves},
  author={Shani, Itamar and Cohen, Gil and Fineberg, Jay},
  journal={Phys. Rev. Lett.},
  volume={104},
  number={18},
  pages={184507},
  year={2010},
  publisher={APS},
  DOI = {10.1103/PhysRevLett.104.184507}
}

@article{milner1991square,
  title={Square patterns and secondary instabilities in driven capillary waves},
  author={Milner, Scott Thomas},
  journal={J. Fluid Mech.},
  volume={225},
  pages={81--100},
  year={1991},
  publisher={Cambridge University Press},
  DOI = {10.1017/S0022112091001970}
}

@article{frumkin2023coupled,
  title={Coupled instabilities drive quasiperiodic order-disorder transitions in Faraday waves},
  author={Frumkin, Valeri and Gokhale, Shreyas},
  journal={Phys. Rev. E},
  volume={108},
  number={1},
  pages={L012601},
  year={2023},
  publisher={APS},
  DOI = {10.1103/PhysRevE.108.L012601}
}

@article{chen1997pattern,
  title={Pattern selection in Faraday waves},
  author={Chen, Peilong and Vi{\~n}als, Jorge},
  journal={Phys. Rev. Lett.},
  volume={79},
  number={14},
  pages={2670},
  year={1997},
  publisher={APS},
  DOI = {10.1103/PhysRevLett.79.2670}
}

@article{bosch1993spatiotemporal,
  title={Spatiotemporal intermittency in the Faraday experiment},
  author={Bosch, Eric and van de Water, Willem},
  journal={Phys. Rev. Lett.},
  volume={70},
  number={22},
  pages={3420},
  year={1993},
  publisher={APS},
  DOI = {10.1103/PhysRevLett.70.3420}
}

@article{zhang1995secondary,
  title={Secondary instabilities and spatiotemporal chaos in parametric surface waves},
  author={Zhang, Wenbin and Vi{\~n}als, Jorge},
  journal={Phys. Rev. Lett.},
  volume={74},
  number={5},
  pages={690},
  year={1995},
  publisher={APS},
  DOI = {10.1103/PhysRevLett.74.690}
}

@article{ezerskii1985random,
  title={Random self-modulation of two-dimensional structures on a liquid surface during parametric excitation},
  author={Ezerskii, AB and Korotin, PI and Rabinovich, MI},
  journal={JETP Lett.},
  volume={41},
  number={4},
  pages={129--131},
  year={1985}
}

@article{zhang1997pattern,
  title={Pattern formation in weakly damped parametric surface waves},
  author={Zhang, Wenbin and Vi{\~n}als, Jorge},
  journal={J. Fluid Mech.},
  volume={336},
  pages={301--330},
  year={1997},
  publisher={Cambridge University Press},
  DOI = {10.1017/S0022112096004764}
}

@article{tufillaro1989order,
  title={Order-disorder transition in capillary ripples},
  author={Tufillaro, NB and Ramshankar, R and Gollub, Jerry P},
  journal={Phys. Rev. Lett.},
  volume={62},
  number={4},
  pages={422},
  year={1989},
  publisher={APS},
  DOI = {10.1103/PhysRevLett.62.422}
}

@article{goodridge1997viscous,
  title={Viscous effects in droplet-ejecting capillary waves},
  author={Goodridge, Christopher L and Shi, W Tao and Hentschel, HGE and Lathrop, Daniel P},
  journal={Phys. Rev. E},
  volume={56},
  number={1},
  pages={472},
  year={1997},
  publisher={APS},
  DOI = {10.1103/PhysRevE.56.472}
}

@article{goodridge1999breaking,
  title={Breaking Faraday waves: critical slowing of droplet ejection rates},
  author={Goodridge, CL and Hentschel, HGE and Lathrop, DP},
  journal={Phys. Rev. Lett.},
  volume={82},
  number={15},
  pages={3062},
  year={1999},
  publisher={APS},
  DOI = {10.1103/PhysRevLett.82.3062}
}

@article{ediger2000spatially,
  title={Spatially heterogeneous dynamics in supercooled liquids},
  author={Ediger, Mark D},
  journal={Annu. Rev. Phys. Chem.},
  volume={51},
  number={1},
  pages={99--128},
  year={2000},
  DOI = {10.1146/annurev.physchem.51.1.99},
  publisher={Annual Reviews 4139 El Camino Way, PO Box 10139, Palo Alto, CA 94303-0139, USA}
}

@article{mehta2010spatial,
  title={Spatial, dynamical and spatiotemporal heterogeneities in granular media},
  author={Mehta, Anita},
  journal={Soft Matter},
  volume={6},
  number={13},
  pages={2875--2883},
  year={2010},
  DOI = {10.1039/B926809J},
  publisher={Royal Society of Chemistry}
}

@article{deseigne2010collective,
  title={Collective motion of vibrated polar disks},
  author={Deseigne, Julien and Dauchot, Olivier and Chat{\'e}, Hugues},
  journal={Phys. Rev. Lett.},
  volume={105},
  number={9},
  pages={098001},
  year={2010},
  doi = {10.1103/PhysRevLett.105.098001},
  publisher={APS}
}

@article{zheng2026topological,
  title={Topological signatures of collective dynamics and turbulent-like energy cascades in apolar active granular matter},
  author={Zheng, Zihan and Jiang, Cunyuan and Chen, Yangrui and Baggioli, Matteo and Zhang, Jie},
  journal={Proc. Nat. Acad. Sci. U. S. A.},
  volume={123},
  number={6},
  pages={e2510873123},
  year={2026},
  doi = {10.1073/pnas.2510873123},
  publisher={National Academy of Sciences}
}

@article{scholz2018rotating,
  title={Rotating robots move collectively and self-organize},
  author={Scholz, Christian and Engel, Michael and P{\"o}schel, Thorsten},
  journal={Nat. Commun.},
  volume={9},
  number={1},
  pages={931},
  year={2018},
  doi = {10.1038/s41467-018-03154-7},
  publisher={Nature Publishing Group UK London}
}

@article{feitosa2002breakdown,
  title={Breakdown of energy equipartition in a 2D binary vibrated granular gas},
  author={Feitosa, Klebert and Menon, Narayanan},
  journal={Phys. Rev. Lett.},
  volume={88},
  number={19},
  pages={198301},
  year={2002},
  doi = {10.1103/PhysRevLett.88.198301},
  publisher={APS}
}

@article{rouyer2000velocity,
  title={Velocity fluctuations in a homogeneous 2D granular gas in steady state},
  author={Rouyer, Florence and Menon, Narayanan},
  journal={Phys. Rev. Lett.},
  volume={85},
  number={17},
  pages={3676},
  year={2000},
  doi = {10.1103/PhysRevLett.85.3676},
  publisher={APS}
}

@article{ghadiri2021control,
  author = {Ghadiri, Mahdi and Krechetnikov, Rouslan},
  title = {Controlling chaos by the system size},
  journal = {Sci. Rep.},
  year = {2021},
  volume = {11},
  number = {1},
  pages={8703},
  month = {APR 22},
  DOI = {10.1038/s41598-021-87233-8},
  Article-Number = {8703},
  ISSN = {2045-2322}
}

@misc{barotta2025macroscopicbrownianmotionchaotic,
      title={Macroscopic Brownian Motion on a Chaotic Fluid Interface}, 
      author={Jack-William Barotta and Caroline M. Barotta and Eli Silver and Daniel M. Harris},
      year={2025},
      eprint={2512.15917},
      archivePrefix={arXiv},
      primaryClass={physics.ed-ph},
      url={https://arxiv.org/abs/2512.15917}, 
}

@misc{sasaki2026constitutiveflowlawhydrogel,
      title={Constitutive flow law for hydrogel granular rafts near the brittle-ductile transition}, 
      author={Yuto Sasaki and Hiroaki Katsuragi},
      year={2026},
      eprint={2602.08217},
      archivePrefix={arXiv},
      primaryClass={cond-mat.soft},
      url={https://arxiv.org/abs/2602.08217}, 
}

@article{kwok1998,
  title={Low-rate dynamic contact angles on polystyrene and the determination of solid surface tensions},
  author={Kwok, DY and Lam, CNC and Li, A and Zhu, K and Wu, R and Neumann, AW},
  journal={Polym. Eng. Sci.},
  volume={38},
  number={10},
  pages={1675--1684},
  year={1998},
  publisher={Wiley Online Library},
  DOI = {10.1002/pen.10338}
}

@article{li2007,
  title={Contact angle of water on polystyrene thin films: Effects of {C}{O}2 environment and film thickness},
  author={Li, Yuan and Pham, Joseph Q and Johnston, Keith P and Green, Peter F},
  journal={Langmuir},
  volume={23},
  number={19},
  pages={9785--9793},
  year={2007},
  publisher={ACS Publications},
  DOI = {10.1021/la0636311}
}

@article{extrand2008,
  title={Contact angles on spherical surfaces},
  author={Extrand, CW and Moon, Sung In},
  journal={Langmuir},
  volume={24},
  number={17},
  pages={9470--9473},
  year={2008},
  publisher={ACS Publications},
  DOI = {10.1021/la801091n}
}

@article{hobson2026structural,
  title={Structural aging of a cohesive and amorphous granular solid under cyclic loading},
  author={Hobson-Rhoades, William and Durian, Douglas J and Fan, Yue and Xiao, Hongyi},
  journal={Soft Matter},
  year={2026},
  volume={22},
  pages={3610--3619},
  publisher={Royal Society of Chemistry},
  DOI = {10.1039/d6sm00108d}
}

@article{kozlowski2019dynamics,
  title={Dynamics of a grain-scale intruder in a two-dimensional granular medium with and without basal friction},
  author={Kozlowski, Ryan and Carlevaro, C Manuel and Daniels, Karen E and Kondic, Lou and Pugnaloni, Luis A and Socolar, Joshua ES and Zheng, Hu and Behringer, Robert P},
  journal={Phys. Rev. E},
  volume={100},
  number={3},
  pages={032905},
  year={2019},
  publisher={APS},
  DOI = {10.1103/PhysRevE.100.032905}
}

@article{miles1994faraday,
  title={Faraday waves: rolls versus squares},
  author={Miles, John},
  journal={J. Fluid Mech.},
  volume={269},
  pages={353--371},
  year={1994},
  publisher={Cambridge University Press},
  DOI = {10.1017/S002211209400159X}
}

@article{van2020propagation,
  title={Propagation and attenuation of mechanical signals in ultrasoft 2D solids},
  author={van Doorn, Jan Maarten and Higler, Ruben and Wegh, Ronald and Fokkink, Remco and Zaccone, Alessio and Sprakel, Joris and van der Gucht, Jasper},
  journal={Sci. Adv.},
  volume={6},
  number={37},
  pages={eaba6601},
  year={2020},
  publisher={American Association for the Advancement of Science},
  DOI = {10.1126/sciadv.aba6601}
}

\end{document}